\documentclass[12pt]{article}
\usepackage{indentfirst} 
\usepackage{amsfonts,amssymb,amsmath}
\usepackage{bm}
\usepackage{cite}
\usepackage{url}
\usepackage{enumitem}
\usepackage{amsthm}

\usepackage{graphicx}
\usepackage[usenames,dvipsnames]{xcolor}
\usepackage{multicol}
\usepackage{MnSymbol}
\usepackage[unicode]{hyperref}
\usepackage{tensor}
\usepackage{stackengine}
\usepackage{tkz-euclide}
\usetikzlibrary{shapes.geometric}
\usepackage[titletoc,title]{appendix}
\usepackage[english]{babel}
\usepackage[utf8]{inputenc}
\usetikzlibrary{arrows}

\AtBeginDocument{}%

\newcommand{\corurl}{red}
\newcommand{\corcite}{ForestGreen}
\newcommand{\corlink}{blue}

\newcommand{\dd}{\mathrm{d}\! \mathrm{I}}

\newcommand\barbelow[1]{\stackunder[1.2pt]{$#1$}{\rule{1.1ex}{.075ex}}}

\hypersetup{linktocpage,colorlinks,urlcolor=\corurl,citecolor=\corcite,linkcolor=\corlink,
pdftitle={Three roads to the Hamiltonian formulation for gravitational theories with boundaries},
pdfauthor={J.F. Barbero, M. Basquens, V. Varo, E.J.S. Villase\~nor}}

\usepackage{xcolor}

\newcommand{\MYhref}[3][blue]{\href{#2}{\color{#1}{#3}}}

\usepackage{titling}
\usepackage{authblk}
\title{Three roads to the geometric constraint formulation of gravitational theories with boundaries}
\author[1,3]{J. Fernando Barbero G.\,}
\author[2]{Marc Basquens\,}
\author[2]{Valle Varo\,}
\author[2,3]{Eduardo J.S. Villase\~nor\,}
\affil[1]{Instituto de Estructura de la Materia, CSIC. Serrano 123, 28006 Madrid, Spain}
\affil[2]{Departamento de Matem\'aticas, Universidad Carlos III de Madrid. Avda.\  de la Universidad 30, 28911 Legan\'es, Spain. Avda.\  de la Universidad 30, 28911 Legan\'es, Spain}
\affil[3]{Grupo de Teor\'{\i}as de Campos y F\'{\i}sica Estad\'{\i}stica. Instituto Gregorio Mill\'an (UC3M). Unidad Asociada al Instituto de Estructura de la Materia, CSIC}
\date{}                     
\begin{document}
	
\maketitle
\date{May 30, 2021}

\vspace*{-12ex}

\begin{abstract}
The Hamiltonian description of mechanical or field models defined by singu\-lar Lagrangians plays a central role in physics.  A number of methods are known for this purpose, the most popular of them being the one developed by Dirac. Here, we discuss other approaches to this problem that rely on the direct use of the equations of motion (and the tangency requirements characteristic of the Gotay, Nester, Hinds method), or are formulated in the tangent bundle of the configuration space. Owing to its interesting relation with general relativity we will use a concrete example as a test bed: an extension of the Pontryagin and Husain-Kucha\v{r} actions to four dimensional manifolds with boundary.
\end{abstract}


\tableofcontents

\medskip

%
%
\section{Introduction and preliminary remarks}{\label{sec_Intro}}

In the present paper, we discuss the Hamiltonian treatment of some field theories with boundaries with applications to gravitational physics, such as the Husain--Kucha\v{r}--Pontryagin model and lower dimensional general relativity. Boundaries play a prominent role in gravitational physics, for instance, they can be used to model black holes with different types of horizons \cite{WaldLR,AK}, to study the asymptotic behavior of solutions to the Einstein's equations \cite{P1, AS, ACL, AshtekarEngleSloan2008} and holography \cite{tHooft:1984kcu,Smolin:1998qp}. The specific reason why we consider the Husain--Kucha\v{r}--Pontryagin model in this context is because, as it has been shown in \cite{HKP}, it has a neat physical interpretation in a four-dimensional manifold with boundary because the boundary theory is the extension of the three-dimensional, Euclidean, general relativity known as the Baekler--Mielke \cite{BM} model. This suggests that it may be possible to find a field theory in five dimensions---with simple enough dynamics---leading to four-dimensional general relativity as the boundary theory. In order to study this model it would be necessary to have the right tools to deal with the presence of the boundary. This is what we aim at providing here. An additional justification to consider the Hamiltonian formulation for this model is its close relationship with the Ashtekar formulation of general relativity both at the Hamiltonian and quantum levels because the phase spaces of both theories coincide, as do most of the constraints (all of them, with the important exception of the scalar constraint). From the point of view of the quantum theory, the kinematical Hilbert space of the Husain--Kucha\v{r} model is, \emph{precisely}, the kinematical Hilbert space of loop quantum gravity. For all these reasons, and given the relevance of boundaries in gravitational physics, we think that it is very important to provide efficient and easy-to-use Hamiltonian methods that are adapted to be used in the presence of boundaries. This is the main purpose of the~paper.\\

The Hamiltonian formulation of field theories defined by singular Lagrangians (among which general relativity is a famed example) has a long history. A turning point in the quest for a systematic treatment of these systems was the introduction by Dirac of his celebrated ``algorithm'' \cite{Dirac} which has been in use ever since. Strictly speaking, Dirac's method as originally conceived, works only for mechanical systems with a finite number of degrees of freedom. Despite the statements made in this regard by Dirac himself \cite[p.26]{Dirac}, the extension of his method to field theories is not immediate. One has to proceed with care because some basic objects are not well defined, in particular the Poisson brackets between canonically conjugate variables. For instance, in the case of a scalar field $\phi$ with canonical momentum $\pi$, it is customary to write
\begin{equation}\label{PB}
\{\phi(x),\pi(y)\}=\delta(x,y)\,,
\end{equation}
and work with this expression. However, Poisson brackets can only be defined for differentiable functions in phase space and are, themselves, differentiable functions. Although the appearance of a Dirac delta seems to be an acceptable departure from smoothness that can be taken care of by resorting to simple tricks like smearing, it is not difficult to come up with models where formal expressions like \eqref{PB} fail to work in a glaring way. Among such models, field theories in bounded regions stand out.

An effective way to avoid the problems that originate from the use of formal expressions, such as \eqref{PB}, is to use the GNH method \cite{GNH1,GNH2,GNH3} or a geometric rephrasing of the original Dirac approach \cite{BDMV}. The crucial element in these alternative methods is to require the Hamiltonian vector fields, whose integral curves define the system's dynamics, to be tangent to the constraint submanifold in phase space. As these tangency conditions may be written and studied without the use of Poisson brackets, many of the actual difficulties found in concrete computations in manifolds with boundaries disappear.
Another source of difficulties (and misunderstandings) when dealing with field theories in bounded regions has to do with the behaviour of the fields at the boundaries and its relationship with the dynamics (see, for instance, \cite{BPV} and references therein).

In the following, we will restrict ourselves to field theories derived from action principles. Let us consider the spacetime $M$ where the field theory is defined to be, unless otherwise stated, (diffeomorphic to) the product of a finite interval of the real line and a 3-dimensional manifold $\Sigma$ with (possibly empty) boundary $\partial\Sigma$, i.e. $M=[t_1,t_2]\times \Sigma$ with $t_1<t_2$. We will often refer to the sets $\Sigma_1:=\{t_1\}\times \Sigma$ and $\Sigma_2:=\{t_2\}\times \Sigma$ as the \emph{lids} and $\partial_LM:=[t_1,t_2]\times \partial \Sigma$ as the \emph{lateral boundary} (see figure 1).

Being the action $S$ a functional on certain space of fields over $M$, not only it is necessary to define the independent fields that will be used to write it (the field space $\mathcal{F}$, often consisting of sections of some tensor bundle), but also to consider their smoothness properties. This is usually done by requiring the fields to live in appropriate functional spaces. As a part of this specification it is possible to introduce boundary conditions.

Actions are usually written as integrals of top forms on the spacetime manifold $M$. In the case of manifolds with boundaries, additional contributions associated with the boundary may also be included (an instance of this is the Gibbons-Hawking-York boundary term in metric gravity). In general, an action will be defined by a Lagrangian pair $(L,\ell)$ of top forms defined on $M$ and its lateral boundary $\partial_L M$.

\begin{figure}[htp]
\begin{center}

\hspace*{-13mm}\begin{tikzpicture}[scale=1.5]
\draw[rounded corners=35pt](0,0)--(0,4);
\draw[rounded corners=35pt](3,0)--(3,4);

\draw[very thin](-1,0)--(-1,4); 
\draw[very thin](-1.1,2)--(-0.9,2); 

\draw[rounded corners=35pt](1.7,0)--(1.7,4);
\draw[->, rounded corners=35pt](1.7,0)--(1.7,2.5);
\draw[gray, dashed](1.7,-0.9)--(1.7,0);
\draw[rounded corners=35pt](0.9,0)--(0.9,4);
\draw[->,rounded corners=35pt](0.9,0)--(0.9,2.5);
\draw[gray, dashed](0.9,-0.9)--(0.9,-0.6);
\draw[gray, dashed](0.9,-0.6)--(0.9,0);

\draw (3,-0.9) arc (0:360:1.5cm and 0.3cm);

\draw [->] (-0.4,-0.9) arc (190:170:8.3cm);

\draw (3,4) arc (0:360:1.5cm and 0.3cm);
\draw [gray, dashed](3,2) arc (0:180:1.5cm and 0.3cm);
\draw (3,2) arc (360:180:1.5cm and 0.3cm);
\draw [gray, dashed](3,0) arc (0:180:1.5cm and 0.3cm);
\draw (3,0) arc (360:180:1.5cm and 0.3cm);

\node (a) at (1.5, 4.7) {$[t_1,t_2]\times \Sigma$};
\node (b) at (3.3, 2) {$\Sigma_{t}$};
\node (c) at (-0.72,.5) {$\jmath_{t}$};
\node (d) at (-1.2, 2.1) {$t$};
\node (e) at (3.3, -0.8) {$\Sigma$};
\node (f) at (-1, 4.2) {$[t_1,t_2]$};
\node (g) at (1.9, 2.5) {$\partial_{t}$};
\node (h) at (1.1, 2.5) {$\partial_{t}$};
\node (i) at (1.3, 3.5) {$\partial_LM$};
\node (j) at (1.1,-0.9) {$p_{1}$};
\node (k) at (1.9,-0.9){$p_{2}$};
\node (l) at (3.3, 4) {$\Sigma_2$};
\node (m) at (3.3, 0) {$\Sigma_1$};

\foreach \point in {(0.9,-0.9),(1.7,-0.9),(0.9,2),(1.7,2)}{
    \fill \point circle (1pt);
}
\end{tikzpicture}
\medskip

\textbf{Figure 1.} Spacetime topology
\end{center}
\end{figure}

Given an action and the values of the fields at the lids $\Sigma_1$ and $\Sigma_2$, the dynamics is obtained by looking for its stationary points. The stationarity conditions will generically consist of equations in the bulk and equations at the boundary. Some comments are in order:

\begin{itemize}

\item[i)] If boundary conditions have been introduced in the definition of $\mathcal{F}$, it is critical to take them into account when deriving the boundary Euler-Lagrange equations. This is so because the variations at the boundary will not be independent but will be constrained by the boundary conditions.

\item[ii)] Even when no boundary contribution is included in the action, there may still exist Euler-Lagrange equations at the boundary in addition to those at the bulk (usually coming from integrations by parts). As a consequence, the issue mentioned in the previous item will still be relevant \cite{Regge:1974zd}.

\medskip

\item[iii)] It is very important to understand that boundary conditions may appear as Euler-Lagrange equations at the boundary even if no such conditions are introduced in $\mathcal{F}$. Also, they do not need to be simple specifications of the values of some fields or their ``spatial'' derivatives but may be \textit{dynamical} (this will happen in the models that we consider here).

\end{itemize}

The purpose of this paper is to explore three different ways to obtain the Hamiltonian formulation of field theories linear in velocities in bounded regions.  This is important in general relativity because the actions used in some relevant approaches (in particular the Hilbert-Palatini or Holst actions) are \textit{precisely} of this type.
The first and second approaches are based on the geometric constraint algorithm: the first one in the cotangent bundle \cite{HKP} and the second in the tangent bundle \cite{Nester1988,carinena1,Lecanda_Roman}.
In a different spirit, the third procedure starts right off from the field equations and quickly arrives at the Hamiltonian formulation \cite{Holstnos}.
These ideas have been known for quite some time, but have not been widely applied when boundaries are present, at least in the context of gravitational theories. As will be shown later, for the type of models discussed in the present paper, the final Hamiltonian description can be made in a phase space which is a submanifold of the configuration space.

A few words on notation. As the basic fields that we will be using are differential forms we will not need to use spacetime indices, however, will use  internal $SO(3)$ indices $i\,,j\,,\ldots=1,2,3$ which may be raised and lowered with the $SO(3)$ invariant metric $\delta_{ij}$ and its inverse $\delta^{ij}$. We will also use the $SO(3)$ volume form $\varepsilon_{ijk}$. If we have a volume form $\mathsf{vol}$ in a manifold $M$ and we have another top form $\alpha$ in $M$ there exists a unique scalar field $\phi$ such that $\alpha=\phi\mathsf{vol}$. We will often denote $\phi=\left(\frac{\alpha}{\mathsf{vol}}\right)$.

%
%
\section{Some basic facts about the Husain-Kucha\v{r} model}{\label{sec_HK}}

The Husain-Kucha\v{r} (HK) model was introduced in \cite{HK} to understand some features of the Ashtekar formulation of general relativity \cite{Asht1,Asht2} (see also \cite{Varadarajan:1999aj}), since their respective Hamiltonian descriptions share the same phase space (the main difference being the absence of the Hamiltonian constraint in the HK model). The action of the HK model reads
\begin{equation}\label{action_HK}
S(e,A)=\int_M\varepsilon_{ijk}e^i\wedge e^j\wedge F^k,
\end{equation}
where here $M$ is a closed, parallelizable (hence, orientable), 4-dimensional manifold,  and $e^i\,,A^i\in\Omega^1(M)$, $i=1,2,3$, are 1-form fields. At each point $p\in M$, the three covectors $e_i(p)$ are required to be linearly independent (this is part of the specification of the configuration space of the model), and
\begin{subequations}\label{defsSO3}
\begin{align}
&De^i:=de^i+\varepsilon^i\,_{jk}A^j\wedge e^k\,,\label{covd}\\
&F^i:=dA^i+\frac{1}{2}\varepsilon^i\,_{jk}A^j\wedge A^k\,,\label{curv}
\end{align}
\end{subequations}
are the $SO(3)$ covariant derivative of the $e^i$ and the curvature of the $SO(3)$ connection $A_i$, respectively. The field equations are:
\begin{subequations}\label{eqsHK}
\begin{align}
&\varepsilon_{ijk}e^j\wedge De^k=0\,,\label{EqDE}\\
&\varepsilon_{ijk}e^j\wedge F^k=0\,.\label{EqF}
\end{align}
\end{subequations}
Structurally, they resemble the Einstein equations derived from the Hilbert-Palatini action. This explains the connection between the HK model and general relativity. Notice, anyway, that in this example the (0,2)-tensor
$\gamma:=e_i\otimes e^i$ is a degenerate metric as we do not have a co-frame but only three independent 1-forms $e_i$.
The degenerate directions of this metric can be easily characterized. If we choose a volume form $\mathsf{vol}$ in $M$ (which is always possible because $M$ is orientable) we can write
\begin{equation}\label{u}
u(\cdot):=\left(\frac{\cdot\wedge \varepsilon_{ijk}e^i\wedge e^j\wedge e^k}{\mathsf{vol}}\right)\,,
\end{equation}
which at each point $p\in M$ is an element of the double dual $T_p^{**}\!M$ of the tangent space $T_pM$. As $T_p^{**}\!M$ is canonically isomorphic to $T_p M$ the previous expression actually defines a vector field $\bm{u}\in\mathfrak{X}(M)$. Since $u(e^i)=0$, then $\gamma(\bm{u},\cdot)=0$, and hence the degenerate directions of the metric are those given by $\bm{u}$. Notice that, if we change the fiducial volume form $\mathsf{vol}$, the direction of the field $\bm{u}$ at each point of $M$ stays the same, although the vector itself will be rescaled.

The field equations \eqref{eqsHK} admit a simple geometric interpretation based on the fact (see Appendix \ref{app_lemma}) that if $e_i$ are three linearly independent frame fields (1-
forms) on a four-dimensional manifold $M$ and $S_i$ are three 1-forms on $M$ satisfying
$\varepsilon_{ijk}e^j\wedge S^k= 0$, then $S_i = 0$. Now, as $\imath_{\bm{u}}e^i=0$, the field equations \eqref{eqsHK} imply
\begin{align*}\label{EqmodifiedU}
&\varepsilon_{ijk}e^j\wedge\imath_{\bm{u}}De^k=0\,,\\
&\varepsilon_{ijk}e^j\wedge\imath_{\bm{u}}F^k=0\,,
\end{align*}
and, hence, $\imath_{\bm{u}}De^k=0$ and $\imath_{\bm{u}}F^k=0$. A straightforward computation then gives
\begin{align*}
&\imath_{\bm{u}}F^i=\imath_{\bm{u}}(\mathrm{d}A^i+\frac{1}{2}\varepsilon^i_{\phantom{i}jk}A^j\wedge A^k)=\pounds_{\bm{u}}A^i-\mathrm{d}A_{u}^i+\varepsilon^i_{\phantom{i}jk}A_{u}^jA^k=\pounds_{\bm{u}}A^i-DA_{u}^i\,,\\
&\imath_{\bm{u}}De^i=\imath_{\bm{u}}(\mathrm{d}e^i+\varepsilon^i_{\phantom{i}jk}A^j\wedge e^k)=\pounds_{\bm{u}}e^i+\varepsilon^i_{\phantom{i}jk}A_{u}^je^k\,,
\end{align*}
where $A_u^i:=\imath_{\bm{u}}A^i$. Hence, we conclude that
\begin{align}
&\pounds_{\bm{u}}A^i=DA_{u}^i\,,\\\label{liedraing}
&\pounds_{\bm{u}}e^i=-\varepsilon^i_{\phantom{i}jk}A_{u}^je^k\,.
\end{align}
The meaning of the dynamics is then clear: the effect of Lie-dragging a solution of the field equations along the direction defined by $\bm{u}$, is just an $SO(3)$ gauge transformation with parameter $A_u^i$. We will see in section \ref{subsec_tangency} how a similar reasoning allows us to extend the previous analysis to an arbitrary vector field instead of $\bm{u}$. From the point of view of the degenerate metric $\gamma$ the interpretation of the dynamics is also clear: it will just be Lie-dragged along the integral curves of the vector field $\bm{u}$.

The physical content of the model is simple to describe: whereas general relativity has two local physical degrees of freedom per point, the Husain-Kucha\v{r} model has three. In both instances they are contained in an $SO(3)$ connection and its canonically conjugate densitized triad, hence, it is a bit surprising that in general relativity the natural variables are non-degenerate 4-metrics whereas it does not seem possible to build these metrics in the HK model. A partial answer to this problem is discussed in \cite{HKPhi}, where it was shown that, by adding an scalar field playing the role of time, it was possible to build non-degenerate 4-dimensional metrics for the HK model. An interesting question in this regard ---which to our knowledge has not been answered yet--- is the characterization of those metrics of the type described in \cite{HKPhi} which also solve the Einstein equations.

The constraint submanifold in phase space for general relativity in Ashtekar variables is a submanifold of the constraint submanifold for the HK model so, from the perspective of the constraints, every solution to the GR constraints is also a solution to the HK ones. Notice, however, that in order to define the GR dynamics in the latter context, it is necessary to introduce the appropriate Hamiltonian vector field. In contrast with this, at the quantum level, the physical Hilbert space of full GR in the Ashtekar formulation is just a subspace of the one corresponding to the HK midel. The problem in this case is finding the appropriate quantum gravitational observables.

As a final comment, we would like to mention the existence of a number of different action principles that also lead to the Husain-Kucha\v{r} model \cite{HK2con, HKPhi, HKBF}. They provide different points of view that can be useful to understand some features of the model and, eventually to learn something about the Hamiltonian formulation of general relativity.

%
%
\section{The generalised Husain-Kucha\v{r}-Pontryagin action}{\label{sec_HKP}}

From now on, let $M$ be a manifold with boundary. Let us consider $e_i, A_i\in\Omega^1(M)$ with $i = 1\,,2\,,3$ as the basic dynamical fields. They are not subject, \emph{a priori}, to any condition other than smoothness and the requirement that the $e^i$ be linearly independent. In particular, we will impose no boundary conditions on them at this point. The generalisation of the Husain-Kucha\v{r}-Pontryagin action given by \cite{BDMV}, reads
\begin{align}
S(e,A)=\int_{M}&\left(\alpha_1\varepsilon_{ijk}e^i\wedge e^j\wedge F^k+\alpha_2De_ i\wedge De^i+\alpha_3 F_i\wedge F^i\right.\hspace*{1.5cm}\label{action}\\
&\hspace*{2.5cm}\left.+\alpha_4\varepsilon_{ijk}De^i\wedge e^j\wedge e^k+\alpha_5 F_i\wedge De^i\right)\,.\nonumber
\end{align}
where $\alpha_1\,\ldots,\alpha_5\in\mathbb{R}$. The field equations are
\begin{subequations}
\label{eom}
\begin{align}
&(\alpha_1-\alpha_2)\varepsilon_{ijk}e^j\wedge F^k=0\,,\label{eq_bulk_1}\\
&(\alpha_1-\alpha_2)\varepsilon_{ijk}e^j\wedge De^k=0\,,\label{eq_bulk_2}\\
&\jmath_\partial^*\big(2\alpha_2De_i+\alpha_5 F_i+\alpha_4\varepsilon_{ijk}e^j\wedge e^k\big)=0\,,\label{eq_boundary_1}\\
&\jmath_\partial^*\big(\alpha_5De_i+2\alpha_3 F_i+\alpha_1\varepsilon_{ijk}e^j\wedge e^k\big)=0\,.\label{eq_boundary_2}
\end{align}
\end{subequations}

As we can see, demanding stationarity of the action gives two sets of necessary conditions: the bulk equations \eqref{eq_bulk_1}, \eqref{eq_bulk_2}, and the boundary equations \eqref{eq_boundary_1}, \eqref{eq_boundary_2}. It is important to emphasize---we will continue to do so throughout the paper---that the stationarity conditions for an action defined in a manifold with boundary will generically consist of these two types of equations.

Before proceeding further we would like to make several comments
\begin{itemize}
\item[i)] Although we will not discuss in any detail functional analytic issues, something needs to be said about the smoothness conditions on the fields and how they are affected by the presence of a boundary. In the interior of $M$ (the bulk), we will require the fields to be ``smooth enough'' so that the field equations make sense there. In order to make sense of the boundary equations it is also natural to add whatever smoothness requirements on the fields are necessary on $\partial M$. An additional smoothness requirement might also be considered: demanding that, when extended to an open smooth manifold $\widehat{M}$ containing $M$ as a submersion, the bulk equations also hold at $\partial M$.

This last point is relatively subtle. On one hand, it appears unnatural from the viewpoint of the action principle since it does not seem necessary to demand such condition for the stationarity of the action. For instance, if we require the function in the bulk to admit a sufficiently smooth extension to $\widehat{M}$, and if the Euler Lagrange equations have a sufficiently nice form, ``their action on the bulk function'' will be smooth and, by continuity, they will also hold at the boundary.
On the other hand, it can be seen as a sensible requirement that can be imposed \emph{a posteriori} to select a subfamily of solutions to the variational equations with good physical properties, or even, appear as consistency requirements for the dynamics. It is also conceivable that a particular choice of smoothness requirements, both in the bulk and at the boundary, suffices to guarantee extendibility in the above sense. For an ordinary variational problem, the treatment of the lateral boundary and the lids may have to be different.
It may happen that the extendibility condition applies only to lateral boundaries and not to the lids.

Some intuition about these questions can be gained by considering, for example, the Laplace equation on a bounded region of the plane and using the real or imaginary parts of complex analytic functions as examples. The last regularity requirement is, at least at face value, the strongest; we will proceed assuming it in the present work. As a last word of caution it should be mentioned that there may be consistency issues between the smoothness requirements in the bulk and at the boundary that we will also sidestep here.
\item[ii)] According to the regularity conditions that we are considering, the bulk equations must also hold when the fields are restricted to the boundary, so there are several sets of boundary equations. The content of these can be conveniently disentangled by either taking their pullback to the boundary and writing them in terms of pullbacks of the dynamical fields or first computing their interior product with the outer unit normal $\bm{\nu}$ and then pulling them back. This procedure mimics one of the methods that we are going to follow in the paper to obtain the Hamiltonian formulation for the model given by the action \eqref{action}.
\item[iii)] If $\alpha_1=\alpha_2$ there are only boundary equations. The dynamics in the bulk is arbitrary. This means that any field configuration with the correct ``boundary dynamics'' provides stationary points for the action. Otherwise the bulk dynamics is that of the Husain-Kucha\v{r} model. From the point of view of the action, this happens because a simple integration by parts of the terms involving $\alpha_1=\alpha_2$ can be used to cancel them, giving just boundary contributions to the action. Notice that the remaining terms can all be written as total derivatives, so that the action in this case is an integral over the boundary which corresponds to an extension of 3-dimensional (Euclidean) general relativity \cite{BM}. At this point it is worthwhile to advance that the Hamiltonian formulation for this theory will be obtained in the following \emph{in the same footing as} the one corresponding to the bulk model.
\item[iv)] If $4\alpha_2\alpha_3-\alpha_5^2\neq0$ the boundary equations tell us that $\jmath_\partial^*F_i$ and $\jmath_\partial^*De_i$ are proportional to $\jmath_\partial^*(\varepsilon_{ijk}e^j\wedge e^k)$ and, hence, the pullbacks of the bulk equations \eqref{eq_bulk_1} and \eqref{eq_bulk_2} automatically hold as can be seen by plugging the expressions for $\jmath_\partial^*F_i$ and $\jmath_\partial^*De_i$ in terms of $\jmath_\partial^*(\varepsilon_{ijk}e^j\wedge e^k)$ into the pullbacks of \eqref{eq_bulk_1} and \eqref{eq_bulk_2} to the boundary. The physical meaning of the specific models obtained at the boundary for other parameter choices are discussed in \cite{HKP}.
\end{itemize}

In the next subsections, we will focus on three different methods of obtaining the first part of the solution of the evolution problem, that is, an expression for the Hamiltonian vector field and a set of necessary constraints. Notice that this does \textit{not} fully solve the problem since further consistency checks may be needed. Since the three approaches will produce the same results, we will defer this final step until section \ref{subsec_tangency}.

\subsection{GNH analysis in the cotangent bundle}\label{subsec_GNH}

The first approach we would like to present is the Hamiltonian formulation, for the model introduced above, using the geometric GNH approach \cite{GNH1,GNH2,GNH3} (a related analysis using a ``geometrized'' version of the traditional Dirac algorithm \cite{Dirac} can be found in \cite{BDMV}). {The main features of this method are:}
\begin{itemize}
    \item {The final Hamiltonian description lives in the primary constraint submanifold in phase space.}
    \item {Dynamical consistency is rephrased as a tangency condition. This has the advantage of altogether avoiding the use of Poisson brackets, which is useful in spacetime manifolds with boundary.}
    \item {Given a Lagrangian (which may come from a suitable 3 + 1 decomposition of an action) the main steps are: (i) the characterization of the primary constraint submanifold from the definition of momenta (fiber derivative), (ii) the definition of the Hamiltonian vector fields in terms of the simplectic form and the exterior derivative in field space of the Hamiltonian and (iii) checking consistency as a tangency requirement.}
\end{itemize}

{To begin with we perform} a $3+1$ decomposition of the action \eqref{action}, we obtain the following Lagrangian \cite{HKP}:

\begin{align}
L(\bm{\mathrm{v}_q})=\int_\Sigma\Big(&(v_A^i-D A_{\mathrm{t}}^i)\wedge(\alpha_5 De_i+2\alpha_3 F_i+\alpha_1\varepsilon_{ijk}e^j\wedge e^k)                   \nonumber\\
&+2\varepsilon_{ijk}e^i_{\mathrm{t}}\wedge e^j \wedge(\alpha_4 De^k+\alpha_1F^k)\label{Lagrangian}\\
&\left.+(v_e^i+\varepsilon^i_{\phantom{i}jk}A^j_{\mathrm{t}}e^k-De_{\mathrm{t}}^i)\wedge(2\alpha_2 De_i+\alpha_5 F_i+\alpha_4\varepsilon_{ilm}e^l\wedge e^m)\right)\,.\nonumber
\end{align}

Here, $A^i\in\Omega^1(\Sigma)$ and $e^i\in\Omega^1(\Sigma)$ are an $SO(3)$ connection and a  non-degenerate frame field on $\Sigma$, respectively. We also have the smooth scalar fields $A_{\mathrm{t}}^i\,,e_{\mathrm{t}}^i\in C^\infty(\Sigma)$. The expressions for the curvature and the covariant derivative that appear in \eqref{Lagrangian} are formally the same as \eqref{curv} and \eqref{covd} but, of course, these objects live now in $\Sigma$. Taken together, the $(A_{\mathrm{t}}^i,A^i,e_{\mathrm{t}}^i,e^i)$ define the configuration space $Q$ for our model  (adding also the requirement that the $e^i$ must be linearly independent) . We will denote the points of $T_qQ$, the tangent space to $Q$ at the point $q=(A_{\mathrm{t}}^i,A^i,e_{\mathrm{t}}^i,e^i)\in Q$, as $\bm{\mathrm{v}_q}$. We will write tangent vectors in the form $\bm{\mathrm{v}_q}=(v_{A_{\mathrm{t}}}^i, v_A^i,v_{e_{\mathrm{t}}}^i, v_e^i)$. As we can see the Lagrangian is a real function in $TQ$.

{The fiber derivative (i.e., the definition of the canonical momenta) associated with a Lagrangian $L$ is a map from the tangent bundle of the configuration space $Q$ to its contangent bundle (phase space)
\[
FL:TQ\rightarrow T^*Q:(q,v)\mapsto (q,p)\,,\,\,p\in T_q^*Q\,,
\]
with
\[
\langle p,w\rangle:=\frac{\mathrm{d}}{\mathrm{d}t}L(q,v+t w)\Big|_{t=0}\,,\,\,v,w\in T_qQ\,.
\]

In the present case this yields}
\begin{align}
\langle FL(\mathrm{\bm{\mathrm{v}_q}}), \bm{\mathrm{w}_q}\rangle=\int_\Sigma &\Big( w_A^i\wedge\big(\alpha_5De_i+2\alpha_3F_i+\alpha_1\varepsilon_{ijk}e^j\wedge e^k\big)\label{fiber_derivative}\\
&+w_e^i\wedge\big(2\alpha_2De_i+\alpha_5F_i+\alpha_4\varepsilon_{ijk}e^j\wedge e^k\big)\Big)\,,\nonumber
\end{align}
so that the canonical momenta   $(\bm{\mathrm{P}}_{A_{\mathrm{t}}}\,,\bm{\mathrm{P}}_{A}\,,\bm{\mathrm{P}}_{e_{\mathrm{t}}}\,,\bm{\mathrm{P}}_{e})$ are {defined by}
\begin{align}\label{momenta}
&\hspace{-0.5cm}\bm{\mathrm{P}}_{A_{\mathrm{t}}}(w_{A_{\mathrm{t}}}^i)&&\hspace*{-4mm}:= \langle FL(\bm{\mathrm{v}_q}),(w_{A_{\mathrm{t}}}^i,0,0,0)\rangle&&\hspace*{-4mm}=0\,,\\
&\hspace{-0.5cm}\bm{\mathrm{P}}_{A}(w_A^i)&&\hspace*{-4mm}:= \langle FL(\bm{\mathrm{v}_q}),(0,w_A^i,0,0)\rangle&&\hspace*{-4mm}=\int_\Sigma w_A^i\wedge(\alpha_5 De_i+2\alpha_3 F_i+\alpha_1 \varepsilon_{ijk}e^j\wedge e^k)\,,\\
&\hspace{-0.5cm}\bm{\mathrm{P}}_{e_{\mathrm{t}}}(w_{e_{\mathrm{t}}}^i)&&\hspace*{-4mm}:= \langle FL(\bm{\mathrm{v}_q}),(0,0,w_{e_{\mathrm{t}}}^i,0)\rangle&&\hspace*{-4mm}=0\,,\\
&\hspace{-0.5cm}\bm{\mathrm{P}}_{e}(w_e^i)&&\hspace*{-4mm}:= \langle FL(\bm{\mathrm{v}_q}),(0,0,0,w_e^i)\rangle&&\hspace*{-4mm}=\int_\Sigma w_e^i\wedge(2\alpha_2 De_i+\alpha_5 F_i+\alpha_4 \varepsilon_{ijk}e^j\wedge e^k)\,.
\end{align}

The fiber derivative is not a diffeomorphism from $TQ$ to $T^*Q$ because it is not onto, hence the dynamical system defined by the action \eqref{action} is singular. The image of $TQ$ under the fiber derivative $FL$ is the so called primary constraint submanifold $\mathfrak{M}_0$ of the phase space $T^*Q$; the dynamics of the system is constrained to $\mathfrak{M}_0$. The Hamiltonian is defined only on this primary constraint submanifold. In the present case it is
\begin{align}
H=\int_\Sigma &\Big(DA_{\mathrm{t}}^i\wedge(\alpha_5De_ i+2\alpha_3F_i+\alpha_1\varepsilon_{ijk}e^j\wedge e^k)-2\varepsilon_{ijk}e_{\mathrm{t}}^ie^j\wedge(\alpha_4De^k+\alpha_1 F^k) \label{Hamiltonian}\\
&\hspace*{2cm}-(\varepsilon^i_{\phantom{i}jk}A^j_{\mathrm{t}}e^k-De^i_{\mathrm{t}})\wedge(2\alpha_2De_ i+\alpha_5F_i+\alpha_4\varepsilon_{imn}e^m\wedge e^n)\Big)\,.\nonumber
\end{align}

It is interesting to notice that it does not depend on the canonical momenta.

Vector fields in phase space will have components
\[
\mathbb{Z}=(Z_{A\mathrm{t}}^i\,,Z_A^i\,, Z_{e\mathrm{t}}^i\,, Z_e^i\,, \bm{\mathrm{Z}}_{\!A\mathrm{t}i}\,, \bm{\mathrm{Z}}_{\!Ai}\,,,\bm{\mathrm{Z}}_{\!e\mathrm{t}i}\,,\bm{\mathrm{Z}}_{\!ei} )\,,
\]
where the boldface letters denote the momenta directions in phase space. Notice that $Z_A^i\,,Z_e^i\in\Omega^1(\Sigma)$ and $Z_{A\mathrm{t}}^i\,,Z_{e\mathrm{t}}^i\in C^\infty(\Sigma)$ and, hence, it makes sense to consider their pullbacks to $\partial\Sigma$.

Acting on vector fields $\mathbb{Z}\,,\mathbb{Y}\in\mathfrak{X}(T^*Q)$ the canonical symplectic form is
\begin{align}
\Omega(\mathbb{Z}, \mathbb{Y})=&\bm{\mathrm{Y}}_{\!A\mathrm{t}i}(Z^i_{A\mathrm{t}})-\bm{\mathrm{Z}}_{\!A\mathrm{t}i}(Y^i_{A\mathrm{t}})
+\bm{\mathrm{Y}}_{\!Ai}(Z^i_{A})-\bm{\mathrm{Z}}_{\!Ai}(Y^i_{A})\label{symplectic}\\
+&\bm{\mathrm{Y}}_{\!e\mathrm{t}i}(Z^i_{e\mathrm{t}})\,\,-\,\,\bm{\mathrm{Z}}_{\!e\mathrm{t}i}(Y^i_{e\mathrm{t}})
\,\,+\,\bm{\mathrm{Y}}_{\!ei}(Z^i_{e})\,\,-\,\bm{\mathrm{Z}}_{\!ei}(Y^i_{e})\,.\nonumber
\end{align}

We have to now obtain the pullback $\omega$ of $\Omega$ to the primary constraint submanifold $\mathfrak{M}_0$. A straightforward computation yields
\begin{align}
\omega(\mathbb{Z},\mathbb{Y})=&\int_\Sigma 2(\alpha_1-\alpha_2)\varepsilon_{ijk}\big(Z_A^i\wedge Y_e^j-Z_e^i\wedge Y_A^j\big)\wedge e^k\label{omega_pullback}\\
-&\int_{\partial\Sigma}\left(\alpha_5(\jmath_\partial^*Z_{ei}\wedge\jmath_\partial^*Y_A^i+\jmath_\partial^*Z_{Ai}\wedge\jmath_\partial^*Y_e^i)
+2\alpha_3\jmath_\partial^*Z_{Ai}\wedge\jmath_\partial^*Y_{A}^{i}+2\alpha_2\jmath_\partial^*Z_{ei}\wedge\jmath_\partial^*Y_{e}^{i}\right)\,.\nonumber
\end{align}

We compute now $\dd H(\mathbb{Y})$
\begin{align}
\hspace*{-.8cm}\dd H(\mathbb{Y})=2(\alpha_1-\alpha_2)\int_\Sigma &\Big(-Y_{A\mathrm{t}}^i\wedge\varepsilon_{ijk} De^j\wedge e^k+Y_{e\mathrm{t}}^i\wedge \varepsilon_{ijk}F^j\wedge e^k\nonumber\\
&\hspace*{.1cm}-Y_A^i\wedge\big(D(\tensor{\varepsilon}{_i_j_k}e_{\mathrm{t}}^{j}e^k)-e_i\wedge (A_{\mathrm{t}j}e^j)\big)\nonumber\\
&\hspace*{.1cm}-Y_e^i\wedge\big(\tensor{\varepsilon}{_i_j_k}F^j e_{\mathrm{t}}^k-D(\tensor{\varepsilon}{_i_j_k}A_{\mathrm{t}}^j)\wedge e^k \big)\Big)\nonumber\\
&\hspace*{-.8cm}+\int_{\partial\Sigma}\jmath_\partial^*\Big(Y_{A\mathrm{t}}^i\big(\alpha_5 De_i+2\alpha_3F_i+\alpha_1\varepsilon_{ijk}e^j\wedge e^k\big)\\
& \hspace*{.7cm}+Y_{e\mathrm{t}}^i\big(2\alpha_2 De_i+\alpha_5F_i+\alpha_4\varepsilon_{ijk}e^j\wedge e^k\big)\nonumber\\
&\hspace*{.6cm}+Y_A^i\wedge\big(2\alpha_3 DA_{\mathrm{t} i}- 2\alpha_1 \tensor{\varepsilon}{_i_j_k} e_{\mathrm{t}}^j e^k-\alpha_5(\tensor{\varepsilon}{_i_j_k}A_{\mathrm{t}}^j e^k-De_{\mathrm{t} i})\big)\nonumber\\
&\hspace*{.7cm}+Y_e^i\wedge\big(\alpha_5DA_{\mathrm{t} i}-2\alpha_4\tensor{\varepsilon}{_i_j_k} e_{\mathrm{t}}^j e^k-2\alpha_2(\tensor{\varepsilon}{_i_j_k} A_{\mathrm{t}}^j e^k-De_{\mathrm{t} i})\big)\Big)\nonumber
\end{align}

By requiring that $\omega(\mathbb{Z},\mathbb{Y})=\dd H(\mathbb{Y})$ for all $\mathbb{Y}\in\mathfrak{X}(\mathfrak{M}_0)$ we obtain {two kinds of equations}:

\begin{itemize}
\item[(1)] Conditions on the components of the vector field $\mathbb{Z}\in \mathfrak{X}(\mathfrak{M}_0)$: There are two types of these associated with the bulk and the boundary, respectively. The bulk conditions are only present if $\alpha_1-\alpha_2\neq 0$ in which case they are
\begin{subequations}\label{Zbulk}
\begin{align}
&\varepsilon_{ijk}Z_A^j\wedge e^k=-\tensor{\varepsilon}{_i_j_k}F^j e_{\mathrm{t}}^k+D(\tensor{\varepsilon}{_i_j_k}A_{\mathrm{t}}^j)\wedge e^k\,,\label{Zbulk1}\\
&\varepsilon_{ijk}Z_e^j\wedge e^k=D\big(\varepsilon_{ijk}e_{\mathrm{t}}^je^k\big)-e_i\wedge(A_{\mathrm{t}j}e^j)\,.\label{Zbulk2}
\end{align}
\end{subequations}
The conditions at the boundary read
\begin{subequations}\label{zbound}
\begin{align}
2\alpha_3\jmath_\partial^*Z_A^i+\alpha_5\jmath_\partial^*Z_e^i&=\jmath_\partial^*\Big(2\alpha_3DA_{\mathrm{t}}^i
-2\alpha_1\varepsilon^i_{\phantom{i}jk}e_{\mathrm{t}}^je^k-\alpha_5(\varepsilon^i_{\phantom{i}jk} A_{\mathrm{t}}^je^k-De_{\mathrm{t}}^i)\Big)\,,\label{Zbound1}\\
\alpha_5\jmath_\partial^*Z_A^i+2\alpha_2\jmath_\partial^*Z_e^i&=\jmath_\partial^*\Big(\alpha_5DA_{\mathrm{t}}^i
-2\alpha_4\varepsilon^i_{\phantom{i}jk}e_{\mathrm{t}}^je^k-2\alpha_2(\varepsilon^i_{\phantom{i}jk} A_{\mathrm{t}}^je^k-De_{\mathrm{t}}^i)\Big)\,.\label{Zbound2}
\end{align}
\end{subequations}

There are no conditions involving $Z^i_{A\mathrm{t}}$ and $Z^i_{e\mathrm{t}}$ neither at the bulk nor at the boundary.

\item[(2)]  Secondary constraints: Again we have constraints associated with the bulk and with the boundary. {They come from the components of $\mathbb{Y}$ in $\dd H(\mathbb{Y})$ that do not appear in $\omega(\mathbb{Z},\mathbb{Y})$ and hence their coefficients must vanish.} The bulk constraints are only present if $\alpha_1-\alpha_2\neq 0$. They are
\begin{subequations}\label{secconstbulk}
\begin{align}
&\varepsilon_{ijk}De^j\wedge e^k=0\label{secconstbulk1} \ ,\\
&\varepsilon_{ijk}F^j\wedge e^k=0\label{secconstbulk2} \ .
\end{align}
\end{subequations}

The boundary constraints are
\begin{subequations}\label{secconstbdry}
\begin{align}
&\jmath_\partial^*\big(\alpha_5De_i+2\alpha_3F_i+\alpha_1\varepsilon_{ijk}e^j\wedge e^k\big)=0\,,\label{secconstbdry1}\\
&\jmath_\partial^*\big(2\alpha_2De_i+\alpha_5F_i+\alpha_4\varepsilon_{ijk}e^j\wedge e^k\big)=0\,.\label{secconstbdry2}
\end{align}
\end{subequations}
\end{itemize}

Although at this point there are still consistency checks to be made---in particular studying the tangency of the Hamiltonian vector fields to the submanifold defined by all the constraints  in phase space---we would now like to draw the attention of the readers to some alternative approaches to the problem.  The tangency analysis will continue in Section \ref{subsec_tangency}.

To conclude this subsection, it is important to highlight the fact that momenta play no role in the Hamiltonian description that we are obtaining. Indeed, the pullback of the symplectic structure to $\mathfrak{M}_0$, the Hamiltonian, the constraints and the Hamiltonian vector fields are all independent of the canonical momenta. {Physically, this means that the dynamics of the momenta is trivial, in the sense already discussed in Section \ref{sec_HK}. Mathematically, this means that the fibers play no role and that the only relevant space is the base configuration space $Q$.} This suggests, for instance, that it is possible to approach the Hamiltonian formulation of the model from the field equations. This issue will be discussed in the next sections.

\subsection{Geometric constraint algorithm in the tangent bundle}\label{subsec_Pepin}

In order to work, the GNH procedure only needs a presymplectic space. In the previous section, this space was $\left( FL(TQ), \omega \right)$. Because the cotangent bundle $T^*Q$ has a canonical symplectic structure, it is a fitting choice for many purposes, in particular there are approaches to quantization that take advantage of the availability of a symplectic or presymplectic form.
On the other hand, it may be interesting to work directly on the tangent bundle of the configuration space where the Lagrangian is defined. This would be natural, for instance, if one wants to apply path integral quantization methods. A possible approach to this would be to import the canonical symplectic form from the phase space via the pullback defined by the fiber derivative $FL$. However, this feels unnatural because it entails going back and forth from $TQ$ to $T^*Q$. It would certainly be desirable to work directly in $TQ$. This can be done as we describe in this section. The main steps of the procedure are the following:
\begin{itemize}
    \item Build a presimplectic form in the tangent bundle of the configuration space $TQ$ from the Lagrangian by using the so called Liouville vector field $\mathbb{V}$.
    \item Define the energy and find the vector fields that give the evolution of the system by solving equation \eqref{eq_lagrangian_Hamilton}.
    \item Impose the second order condition (necessary to guarantee the equivalence of the dynamics with that given by the Euler-Lagrange equations.)
\end{itemize}

Whilst the tangent space does not have a canonical symplectic structure, there are canonical structures in the \emph{double tangent} that can be used to introduce suitable symplectic structures in the tangent bundle of the configuration space $TQ$ once a Lagrangian ---a real function in $TQ$--- is chosen.

If the Lagrangian is singular, we will obtain a presymplectic space in its own right in which the GNH procedure can be applied.
One could raise the issue that this structure is not canonical because it depends on the choice of Lagrangian, however this is not really a problem, in fact, this also happens in the Hamiltonian setting where both the primary constraint manifold and the Hamiltonian are obtained from a choice of Lagrangian which encodes the physics of the system (remember that the primary constraint submanifold in phase space is the image of $TQ$ under the fiber derivative $FL$ defined by the Lagrangian). In fact, under general conditions which hold, in practice, for all interesting physical theories, both formulations are equivalent \cite{GN1}.

The almost tangent structure (or vertical endomorphism) on $TQ$ \cite{GN1, GN2, Crampin_1983} is the vector-valued 1-form $J: TTQ \longrightarrow TTQ$ defined by
\begin{align*}
    J := \xi \circ T\pi_Q \ ,
\end{align*}
where $\xi_y (w) := \frac{d}{d\epsilon} (y+\epsilon w)\rvert_{\epsilon = 0}$ is the canonical lift of $w \in TQ$ to $T_yTQ$ and $\pi_Q: TQ \longrightarrow Q$ is the bundle projection.
It is easy to see that $J^2 = 0$. We define the vertical subspace of $TTQ$ as $V(TQ) = \text{Im }J = \text{ker } J = \text{ker }T\pi_Q$ whose elements are called vertical vectors.
This induces a derivation of rank 0 on differential forms on $TQ$
\begin{align*}
    \imath_J \alpha (X_1, ..., X_p) = \sum_{i=1}^p \alpha(X_1, ... JX_i, ..., X_p) \ ,
\end{align*}
and the vertical derivative
\begin{align*}
    \dd_J := \imath_J \dd - \dd \imath_J \ ,
\end{align*}
such that $\dd_J^2 = 0$.

The Liouville vector field is defined as $\mathbb{V}_y := \xi_y(y)$.
In a natural bundle chart
\begin{align*}
    &J(q, v, \dot{q}, \dot{v}) = (q, v, 0, \dot{q}) \ ,\\
    &\mathbb{V}_{(q, v)} = (0, v) \ .
\end{align*}

With all this, we can build the presymplectic structure $(TQ, \omega_L, \dd E_L)$ associated with the pair $(TQ, L)$, where
\begin{align*}
    \omega_L &= -\dd \dd_J L \ ,\\
    E_L &= \imath_\mathbb{V} \dd L - L \ .
\end{align*}
Indeed, note that the almost tangent structure is canonical to $TQ$ and the Lagrangian was the only other relevant element in the construction.
The Hamiltonian equation in this (pre)symplectic space thus becomes
\begin{align}
\label{eq_lagrangian_Hamilton}
    \imath_\mathbb{Z} \omega_L = \dd E_L \ ,
\end{align}
also referred to as the (pre)symplectic Lagrangian equation.
This equation gives the evolution of the system, but it is very important to realise that if $\omega_L$ is presymplectic, in general, the integral curves of the Lagrangian vector field $\mathbb{Z}$ \textit{are not} a solution of the Euler-Lagrange equations.
The obstruction is that these integral curves in $TQ$ may not be canonical lifts of curves in $Q$, in which case they can not be tied to the variational principle.
To recover equivalence with the Euler-Lagrange equations one must additionally impose the so-called second order condition \cite{Nester1988, GN1}
\begin{align}
\label{eq_second_order_condition}
    J\mathbb{Z}=\mathbb{V} \ ,
\end{align}
which is equivalent to $T\pi_Q (\mathbb{Z})= \pi_{TQ} (\mathbb{Z})$. Then, the stationary points of the action are given by vector fields simultaneously satisfying \eqref{eq_lagrangian_Hamilton} and \eqref{eq_second_order_condition}.

In \cite{GN2} it was proved that there exists a submanifold with an unique vector field solving \eqref{eq_lagrangian_Hamilton} \textit{and} \eqref{eq_second_order_condition}. An algorithmic procedure to obtain such maximal submanifold was later given in \cite{Lecanda_Roman}, we summarize it here:
\begin{itemize}
    \item[i)] A solution to \eqref{eq_lagrangian_Hamilton} exists at $x\in TQ$ if $\left( \dd E_L \right)_x$ is in the image of $\left(\omega_L\right)_x$. This condition can be seen to be equivalent to $\imath_{\mathbb{X}} \left( \dd E_L\right)_x = 0,$ for all $\mathbb{X} \in \left(\text{ker }\omega_L\right)_x$; this is refered to as the \emph{dynamical constraint}. Let $P_1$ be the submanifold where the dynamical constraint is satisfied.
    Note that if $\mathbb{Z}$ is a solution of \eqref{eq_lagrangian_Hamilton}, then $\mathbb{Z} + \mathbb{Y}$ for $\mathbb{Y} \in \text{ker } \omega$ is also a solution.
    \item[ii)] In $P_1$, solutions are guaranteed to satisfy \eqref{eq_lagrangian_Hamilton} but they need not satisfy \eqref{eq_second_order_condition}. Since the solutions will have the form $\mathbb{Z}_0 + \mathbb{Y}$ for $\mathbb{Y} \in \text{ker } \omega_L$, we have some freedom to choose $\mathbb{Y}$ in such a way that $\mathbb{Z}_0 + \mathbb{Y}$ satisfies \eqref{eq_second_order_condition}.
    This can be done in a submanifold $S_1$ of $P_1$ satisfying the condition $\imath_\mathbb{X} \imath_\mathbb{Y}  \left(\omega_L \right)_x = 0$, for all $\mathbb{X} \text{ such that } J\mathbb{X} \in V_x(TQ) \cap \left(\text{ker }\omega_L \right)_x$.
    This is called the \emph{non-dynamical constraint}.
    Note that if $\mathbb{Z} + \mathbb{Y}$ is a solution to both \eqref{eq_lagrangian_Hamilton} and \eqref{eq_second_order_condition}, then $\mathbb{Z} + \mathbb{Y} + \mathbb{W}$ for $\mathbb{W} \in V(TQ) \cap \text{ker }\omega_L$ is also a solution.
    \item[iii)] In $S_1$, solutions to both \eqref{eq_lagrangian_Hamilton} and \eqref{eq_second_order_condition} exist, however they are not tangent to $S_1$ in general.
    Since we still have the freedom to choose $\mathbb{W}  \in V(TQ) \cap \text{ker }\omega_L$, we can take it in such a way that the resulting solution \emph{is} tangent to $S_1$ in a (perhaps smaller) submanifold $S_2$.
    Again, the chosen solution may not be tangent to $S_2$, so we need to iterate this last step until no further constraints crop up.
\end{itemize}

We will apply this method in the case of Lagrangians linear in velocities ---first discussed in \cite{carinena1}--- such as the model \eqref{Lagrangian} that we are studying here.
Such Lagrangians $L \in \mathcal{C}^\infty(TQ)$ are fully characterized by a function $h \in \mathcal{C}^\infty(Q)$ and a 1-form $\mu \in \Omega^1 (Q)$. They can be written as
\begin{align}
\label{eq_linear_lagrangian}
    L = \hat{\mu} + \pi_Q^*h \ ,
\end{align}
where $\hat{\mu}(q, v) := \mu_q (v) \in \mathcal{C}^\infty (TQ)$.
As per the algorithm, the points $x \in TQ$ where \eqref{eq_lagrangian_Hamilton} can be solved are determined by the constraints
\begin{align}
\label{eq_lagrangian_first_submanifold}
\imath_\mathbb{X}(\dd E_L)_x = 0, \ \forall \mathbb{X} \in \text{ker } (\omega_L)_x \ .
\end{align}
It can be shown that for vertical vectors
\begin{align}
\label{eq_vertical_property}
    \imath_\mathbb{X} \dd E_L = 0, \ \forall \mathbb{X} \in  V(TQ) \cap \text{ker }\omega_L \ ,
\end{align}
so that equation \eqref{eq_lagrangian_first_submanifold} actually only imposes conditions on the horizontal vectors.
In the linear-in-velocities case, it is easy to derive the relations
\begin{align*}
    &\imath_\mathbb{V} \dd L = \hat{\mu} \ ,\\
    &\dd E_L = - \pi_Q^* \dd h \ , \\
    &\dd _J \dd \pi_Q^* h = 0 \ , \\
    &\dd  \dd _J \hat{\mu} = \pi_Q^* \dd  \mu \ , \\
    &\omega_L = -\pi_Q^* \dd  \mu \ .
\end{align*}
Since both $\dd E_L$ and $\omega_L$ are pullbacks of objects in $Q$ and vertical vectors do not generate additional restrictions because of \eqref{eq_vertical_property}, we can write the condition \eqref{eq_lagrangian_first_submanifold} in $Q$ as
\begin{align}
\label{eq_lagrangian_first_submanifold_simplified}
    \imath_X(\dd h)_x =0 \ , \forall X \in \text{ker } (\dd \mu)_x \ ,
\end{align}
with $x \in Q$ and $X \in \mathfrak{X}(Q)$.
Note that this means that the dynamical constraints are functions in $Q$ and do not involve the velocities.
The next step consists in finding the points where \eqref{eq_second_order_condition} holds, which are determined by
\begin{align*}
    \imath_\mathbb{X} \imath_\mathbb{Y}  \left(\omega_L \right)_x = 0 \ , \ \forall \mathbb{X} \text{ such that } J\mathbb{X} \in V_x(TQ) \cap \left(\text{ker }\omega_L \right)_x \ , \ J(\mathbb{Z} + \mathbb{Y}) = \mathbb{V} \ .
\end{align*}
The condition for $\mathbb{X}$ is trivial: a vector $\mathbb{W} \in \text{ker }\omega_L$ is such that $W_q \in \text{ker }\dd  \mu$, but since $\mathbb{W} = J \mathbb{X}$, this means that $W_q = 0$, hence all vectors in $\mathfrak{X}(TQ)$ satisfy this condition.
So we just have to demand that there exists a vector $\mathbb{Y} \in \text{ker } \dd \mu$ such that $\mathbb{Z} + \mathbb{Y}$ satisfies the second order condition, or equivalently, that at $x\in TQ$ there exists
\begin{align}
\label{linear_soc}
    Y_q = v - Z_q \ , \text{ and } \ Y_q \in \text{ker } (\dd \mu)_x \ .
\end{align}
Using of the explicit form of \eqref{eq_linear_lagrangian} to rewrite the Hamiltonian equation \eqref{eq_lagrangian_Hamilton} one concludes that the Lagrangian vector field is given by
\begin{align*}
    (D_qD_v \hat{\mu} - D_vD_q \hat{\mu}) \cdot Z_q = D_q h \ ,\\
    Z_v \text{ arbitrary} \ .
\end{align*}
The main result of this analysis is precisely that, when the Lagrangian is linear in the velocities, these do not play any role in either the constraints or the evolution and everything happens in the base space $Q$, making the system particularly easy to analyze.

Turning now to the particular case of the Lagrangian \eqref{Lagrangian} we have
\begin{align*}
\hat{\mu} = \int_\Sigma &\Big(  v_A^i\wedge (\alpha_5 De_i +2\alpha_3 F_i+ \alpha_1\varepsilon_{ijk}e^j\wedge e^k) +  v_e^i\wedge (2\alpha_2 De_i+\alpha_5 F_i+\alpha_4\varepsilon_{ilm}e^l\wedge e^m)\Big) \ ,\\
 h  = \int_\Sigma &\Big( - DA_{\mathrm{t}}^i\wedge(\alpha_5 De_i +2\alpha_3 F_i+ \alpha_1\varepsilon_{ijk}e^j\wedge e^k) +2\varepsilon_{ijk}e^i_{\mathrm{t}} \wedge e^j\wedge(\alpha_4 De^k+\alpha_1F^k)+\\
& +(\varepsilon^i_{\phantom{i}jk}A^j_{\mathrm{t}}e^k-De_{\mathrm{t}}^i)\wedge(2\alpha_2 De_i+\alpha_5 F_i+\alpha_4\varepsilon_{ilm}e^l\wedge e^m)\Big) \ ,
\end{align*}
so that the constraints \eqref{eq_lagrangian_first_submanifold_simplified} are
\begin{align*}
\varepsilon_{ijk}  De^j\wedge e^k = 0 \ ,\\
\varepsilon_{ijk} e^j\wedge F^k = 0 \ ,\\
\jmath_\partial^*\left( \alpha_5 De_i +2\alpha_3 F_i+ \alpha_1\varepsilon_{ijk}e^j\wedge e^k \right) = 0 \ ,\\
 \jmath_\partial^*\left(2\alpha_2 De_i+\alpha_5 F_i+\alpha_4\varepsilon_{ilm}e^l\wedge e^m \right) = 0 \ ,
\end{align*}
and the Lagrangian vector field is determined by
\begin{align*}
\varepsilon_{ijk} Z^j_{A} \wedge e^k =  \varepsilon_{ijk} \left( DA^j_{t} \wedge e^k + e^j_{t} F^k \right) \ ,\\
\varepsilon_{ijk} Z^j_{e} \wedge e^k = \varepsilon_{ijk} D(e^j_{t} e^k) - A^j_{t} e_i\wedge e_j \ ,\\
\jmath_\partial^* \left( 2 \alpha_2 Z_{e_i}+  \alpha_5 Z_{A_i} \right) = \jmath_\partial^* \left( \alpha_5  DA_{t_i} - 2 \alpha_4 \varepsilon_{ijk} e^j_{t} e^k  - 2 \alpha_2 \left( \varepsilon_{ijk} A^j_{t} e^k - De_{t_i} \right)\right) \ ,\\
\jmath_\partial^* \left( \alpha_5 Z_{e_i} +2 \alpha_3 Z_{A_i} \right)= \jmath_\partial^* \left(2 \alpha_3 DA_{t_i} - 2\alpha_1 \varepsilon_{ijk} e^j_{t} e^k - \alpha_5 \left( \varepsilon_{ijk} A^j_{t} e^k - De_{t_i} \right)\right) ,
\end{align*}
with $Z_{e_t}, Z_{A_t}, Z_{v_A}, Z_{v_e}, Z_{v_{A_t}}, Z_{v_{e_t}}$ arbitrary.
Since $\text{ker }d\mu$ consists of the vectors with $Y_A= Y_e=0$, \eqref{linear_soc} can only be satisfied in the points of $TQ$ where
\begin{align*}
    Z_A^i = v_A^i \ , \ Z_e^i = v_e^i \ .
\end{align*}
The constraints and equations for the Lagrangian vector field are the same that we obtained in section \ref{subsec_GNH} using the GNH algorithm. The only exception is the constraints introduced by the second order condition, which is an additional requirement in the tangent bundle and does not appear in the cotangent bundle. The final tangency step will be studied in section \ref{subsec_tangency}.

\subsection{The field equations approach}\label{subsec_field_eqs}

A comparison of the field equations for the model that we are considering here with the constraints obtained by using any of the previous approaches shows that they are \emph{structurally identical}, despite the fact that they are defined on manifolds of different dimensions (the spacetime $M$ and the spatial manifold $\Sigma$). This is not always the case as attested, for instance, by the Einstein equations in metric variables and the constraints in the ADM formulation. The reason for the nice behaviour that we find in our example is easy to understand: in the present case the fields are differential forms and the field equations are written in terms of natural operations such as the exterior derivative and the exterior differential. These operations interact in a very simple and natural way with pullbacks (in particular, by $\jmath_t^*$) and, hence, it is straightforward to obtain \emph{necessary conditions} from the field equations in the form of constraints. In the following we take advantage of this idea to also include dynamics. The main steps of the procedure that we describe in this section are the following:
\begin{itemize}
    \item Pullback the field equations to $\Sigma_t$ by using $\jmath_t^*$ to obtain constraints. To this end it will be useful to first define \emph{adapted fields} by using the fact that $M=[t_1,t_2]\times \Sigma$.
    \item Compute the interior product of the field equations with the vector field $\partial_t$ canonically defined by the decomposition $M=[t_1,t_2]\times \Sigma$ in terms of the objects introduced in the previous step. Then pull this back to $\Sigma_t$.
    \item Write the previous result in terms of time derivatives of the fields and introduce in this way the vector fields that define the evolution of the system.
\end{itemize}

To begin with, recall that $M = I \times \Sigma$, with $I=[t_1, t_2]$, admits a foliation by the hypersurfaces $\Sigma_t$ and the inclusion $\jmath_t : \Sigma \longrightarrow M$ with $\jmath_t ( \Sigma) = \Sigma_t$. As usual, we will denote both the projection on the first argument and its elements by $t$. The vector field in $M$ tangent to the curves $t \mapsto (t, p)$ is denoted by $\partial_t$ and it satisfies
\begin{align*}
    \imath_{\partial_t} dt = 1 \ , \  j_t^* dt = 0 \ .
\end{align*}
Note that a differential form $\alpha \in \Omega^p(M)$ can be \textit{adapted} to the foliation by the decomposition
\begin{align*}
    \alpha = dt \wedge \alpha_t + \barbelow{\alpha} \ ,
\end{align*}
where $\alpha_t := \imath_{\partial_t} \alpha$ and $\barbelow{\alpha} := \imath_{\partial_t} \left( dt \wedge \alpha \right)$. We will call $\alpha_t$ and $\barbelow{\alpha}$ the \emph{adapted components} of $\alpha$.


Consider the family of functions  $\tilde{S}_I: \tilde{Q} \rightarrow \mathbb{R}$ with $I$ an interval of $\mathbb{R}$. They are actions in some configuration space $\tilde{Q}$
\begin{align}
\label{action_f_point}
\tilde{S}_I(\tilde{q}) = \int_{I\times\Sigma} \mathcal{L}\,,
\end{align}
where $\mathcal{L}$ is a top form in $M$ which depends on the objects in $\tilde{Q}$.  The action \eqref{action} is of this form.
One can also rewrite it as a function $S_I: \mathcal{P}(Q) \rightarrow \mathbb{R}$ defined on a space of curves in a different configuration space $Q$ by writing
\begin{align}
\label{action_f_curve}
S_I(\gamma) = \int_{I} dt \int_{\Sigma} \jmath^*_t i_{\partial_t} \mathcal{L} \ ,
\end{align}
where $\gamma$ is constructed with the adapted fields $\jmath^*_t \imath_{\partial_t} \tilde{q}$, $\jmath^*_t \tilde{q} \in \mathcal{P}(Q)$ for each of the configuration variables $\tilde{q}$ in $\tilde{Q}$.
The two formulations are equivalent and the critical points of \eqref{action_f_point} are in one-to-one correspondence with stationary curves of \eqref{action_f_curve}.

It was shown by Nester \cite{Nester1988} that the Euler-Lagrange equations can be written in the invariant form
\begin{align}
\label{eq_Nester}
    \imath_\mathbb{Z} \dd \dd_JL - \dd E_L = 0 \ ,
\end{align}
and that if $\gamma \subset Q$ is a curve solution to the variational principle, the vector field $\mathbb{Z}$ along $\dot{\gamma} \subset TQ$ given by $\mathbb{Z}_{\dot{\gamma}(t)} = \ddot{\gamma}_{\dot{\gamma}(t)} := (\dot{\gamma}_{\gamma(t)}, \ddot{\gamma}_{\gamma(t)})$ solves \eqref{eq_Nester}.
Then, for each $\tilde{q} \in \tilde{Q}$ which is a critical point of \eqref{action_f_point} there is a curve $\gamma$ in $Q$ ---which can be described in terms of the adapted fields--- such that $\mathbb{Z} = \ddot{\gamma}$ solves the Euler-Lagrange equations. Notice that, by construction, $v = Z_q$.
However, a vector field $\mathbb{Z} \in \mathfrak{X}(TQ)$ satisfying \eqref{eq_Nester} will not generally come from a curve $\gamma$ corresponding to a solution of the Euler-Lagrange equations, because the integral curve it generates may not be the canonical lift of a curve in $Q$. This is true only if we additionally impose the second order condition \eqref{eq_second_order_condition}.

According to the previous discussion, if $\Psi$ is an equation of motion produced by variations, the equations
\begin{align}
\begin{split}
\label{eom_method}
    \jmath_t^*\Psi = 0 \ ,\\
    \jmath_t^* \imath_{\partial_t} \Psi = 0 \ ,
\end{split}
\end{align}
give the solution to the symplectic Lagrangian equation \eqref{eq_lagrangian_Hamilton}.
Notice that \eqref{eom_method} will require us to impose the constraints on the points of $Q$ and fix the velocities, which via the procedure explained above are identified with the $q$ components of the Lagrangian vector field. Notice that, \emph{a priori} the resulting vector field need not necessarily be tangent to the submanifold defined by the constraints.

We will now apply this approach to the present problem.
A simple way to do it is by splitting every object in the equations in its adapted components, since then \eqref{eom_method} become trivial. We will (temporarily) denote the fields adapted to $\Sigma$ with a bar to emphasize this and make the computations more transparent.

Let us define the adapted differential as $\barbelow{d} \alpha = \imath_{\partial_t} \left( dt \wedge d\alpha \right)$ which can only act on $\alpha$ adapted to the foliation. Then, the exterior derivative decomposes as
\begin{align*}
    d \alpha = dt \wedge (\pounds_{\partial_t} \barbelow{\alpha} -\barbelow{d} \alpha_t ) + \barbelow{d} \barbelow{\alpha} \ .
\end{align*}
We can also define the covariant derivative on each leaf for $\alpha^i$ adapted to the foliation as
\begin{align*}
    \barbelow{D} \alpha^i = \barbelow{d} \alpha^i + \tensor{\varepsilon}{^i_j_k} \barbelow{A}^j \alpha^k \ ,
\end{align*}
and we will write $\barbelow{F}^i = \barbelow{D} \barbelow{A}^i$ the curvature of $\Sigma$.

As a first step, it is useful to have the decompositions of the covariant derivatives
\begin{align*}
    F^i = \barbelow{D} \barbelow{A}^i + dt \wedge \left( v^i_{\scalebox{0.7}{\ensuremath{\barbelow{A}}}} - \barbelow{D} A_t^i \right)\ ,\\
    De^i = \barbelow{D} \barbelow{e}^i + dt \wedge \left( v^i_{\scalebox{0.7}{\ensuremath{\barbelow{e}}}} + \tensor{\varepsilon}{^i_j_k} A_t^j \barbelow{e}^k - \barbelow{D} e_t^i \right) \ .
\end{align*}
Notice that these expressions depend on the velocities. This is so because of the Lie derivatives that appear in the decomposition of the exterior derivative (since the velocity is precisely the derivative in $t$).

Using these ingredients, the equations of motion \eqref{eom} decompose as
\begin{align*}
    \left( \alpha_1 - \alpha_2 \right) \varepsilon_{ijk}  \barbelow{e}^j \wedge \barbelow{F}^k + dt \wedge \left( \alpha_1 - \alpha_2 \right) \varepsilon_{ijk} \left( e_t^j \barbelow{F}^k - \barbelow{e}^j \wedge \left( v^k_{\scalebox{0.7}{\ensuremath{\barbelow{A}}}} - \barbelow{D} A_t^k\right) \right) &= 0 \ , \\
    \left( \alpha_1 - \alpha_2 \right) \varepsilon_{ijk}  \barbelow{D}\barbelow{e}^j \wedge \barbelow{e}^k + dt \wedge \left( \alpha_1 - \alpha_2 \right) \varepsilon_{ijk}  \left( \left( v^j_{\scalebox{0.7}{\ensuremath{\barbelow{e}}}} + \varepsilon^j_{\phantom{j}ab}A_t^a \barbelow{e}^b - \barbelow{D} e_t^j \right) \wedge \barbelow{e}^k - e_t^j \barbelow{D} \barbelow{e}^k \right) &= 0 \ , \\
    \jmath_\partial^*\left[ \alpha_5 \barbelow{F}^i+ \alpha_4 \tensor{\varepsilon}{^i_j_k} \barbelow{e}^j \wedge \barbelow{e}^k +  2 \alpha_2 \barbelow{D}\barbelow{e^i}\right. + \hspace{.5\textwidth} & \\
    + \left. dt \wedge \left( \alpha_5 \left(v^i_{\scalebox{0.7}{\ensuremath{\barbelow{A}}}} - \barbelow{D} A_t^i \right) +  2\alpha_4 \tensor{\varepsilon}{^i_j_k} e_t^j \barbelow{e}^k+ 2 \alpha_2 \left( v^i_{\scalebox{0.7}{\ensuremath{\barbelow{e}}}} + \tensor{\varepsilon}{^i_j_k} A_t^j \barbelow{e}^k - \barbelow{D} e_t^i \right)  \right) \right] &= 0 \ , \\
    \jmath_\partial^*\left[   2\alpha_3 \barbelow{F}^i + \alpha_1 \tensor{\varepsilon}{^i_j_k} \barbelow{e}^j \wedge \barbelow{e}^k +\alpha_5 \barbelow{D}\barbelow{e^i}\right. + \hspace{.5\textwidth} &\\
    + \left. dt \wedge \left( 2 \alpha_3 \left(v^i_{\scalebox{0.7}{\ensuremath{\barbelow{A}}}} - \barbelow{D} A_t^i \right)  + 2\alpha_1 \tensor{\varepsilon}{^i_j_k} e_t^j \barbelow{e}^k\right) + \alpha_5 \left( v^i_{\scalebox{0.7}{\ensuremath{\barbelow{e}}}} + \tensor{\varepsilon}{^i_j_k} A_t^j \barbelow{e}^k - \barbelow{D} e_t^i \right) \right] &= 0 \ .
\end{align*}

For simplicity, we will drop the bars henceforth.
Since $\mathbb{Z}$ satisfies the second order condition, we can identify the velocities with the corresponding components of the evolution vector field.
Now, extracting both the tangential and transverse components of the equations of motion we obtain the following set of equations:
\begin{align*}
     \left( \alpha_1 - \alpha_2 \right) \varepsilon_{ijk}  e^j \wedge F^k = 0 \ , \\
     \left( \alpha_1 - \alpha_2 \right) \varepsilon_{ijk} \left( e_t^j F^k - e^j \wedge \left( Z_A^k - D A_t^k\right) \right) = 0 \ , \\
     \left( \alpha_1 - \alpha_2 \right) \varepsilon_{ijk}  De^j \wedge e^k = 0 \ , \\
     \left( \alpha_1 - \alpha_2 \right) \varepsilon_{ijk}  \left( \left(  Z_e^k  + \varepsilon^j_{\phantom{j}ab}A_t^a e^b - D e_t^j \right) \wedge e^k - e_t^j D e^k \right) = 0 \ , \\
     \jmath_\partial^*\left[  \alpha_5 F^i+ \alpha_4 \tensor{\varepsilon}{^i_j_k} e^j \wedge e^k+2 \alpha_2 De^i \right] = 0 \ , \\
     \jmath_\partial^*\left[ \alpha_5 \left( Z_A^i- D A_t^i \right) +2\alpha_4 \tensor{\varepsilon}{^i_j_k} e_t^j e^k+ 2 \alpha_2 \left(  Z_e^i + \tensor{\varepsilon}{^i_j_k} A_t^j e^k - D e_t^i \right)  \right] = 0 \ , \\
     \jmath_\partial^*\left[ 2\alpha_3 F^i + \alpha_1 \tensor{\varepsilon}{^i_j_k} e^j \wedge e^k + \alpha_5 De^i  \right] = 0 \ , \\
    \jmath_\partial^*\left[ 2 \alpha_3 \left( Z_A^i - D A_t^i \right)  + 2\alpha_1 \tensor{\varepsilon}{^i_j_k} e_t^j e^k + \alpha_5 \left( Z_e^i + \tensor{\varepsilon}{^i_j_k} A_t^j e^k - D e_t^i \right) \right] = 0 \ .
\end{align*}
Some of them,  which come from the tangent part of the equations, are conditions involving only the points in $Q$: these are constraints. The remaining ones, coming from the transverse part of the equations, involve the components of the vector field $\mathbb{Z}$. These define the evolution of the system. As can be readily seen, both the constraints and the evolution equations are \emph{precisely} the same ones that we have found with the other two methods.

\subsection{Final consistency analysis}\label{subsec_tangency}

Here we give a detailed account and extend the analysis presented in \cite{HKP}. As we have shown the three different methods discussed above give the same result in the form of constraints and equations for the components of the Hamiltonian vector field $\mathbb{Z}$ associated both with the bulk and the boundary. From this point on the consistency analysis that will eventually lead to the final Hamiltonian formulation for the model considered here is the same for the three cases. The main issue to be checked is the tangency of the Hamiltonian vector field $\mathbb{Z}$ to the submanifold defined by all the constraints.

The equations for $\mathbb{Z}$ can be solved in a straightforward way without introducing new secondary constraints and will depend only on the configuration variables \cite{HKP}. The bulk components can be easily obtained by solving the equations \eqref{Zbulk} (see \cite{HKP} and \cite{Holstnos}). To this end it is convenient to rewrite them in the form
\begin{subequations}\label{Zbulk_new}
\begin{align}
&\varepsilon_{ijk}e^j\wedge (Z_A^k-DA_{\mathrm{t}}^k)=\varepsilon_{ijk}e_{\mathrm{t}}^jF^k \label{Zbulk1_new}\\
&\varepsilon_{ijk}e^j\wedge (Z_e^k-De_{\mathrm{t}}^k+\varepsilon^k_{\phantom{k}lm}A_{\mathrm{t}}^le^m)=\varepsilon_{ijk}e_{\mathrm{t}}^jDe^k\,.\label{Zbulk2_new}
\end{align}
\end{subequations}
The solutions are
\begin{subequations}\label{solZ_bulk}
\begin{align}
Z_A^i&=DA_{\mathrm{t}}^i+\varepsilon_{jkl}\left(\frac{e^i\wedge F^j}{\omega}\right)e_{\mathrm{t}}^ke^l\,,\label{solZA_bulk}\\
Z_e^i&=De_{\mathrm{t}}^i-\varepsilon^i_{\phantom{i}lm}A_{\mathrm{t}}^le^m+\varepsilon_{jkl}\left(\frac{e^i\wedge De^j}{\omega}\right)e_{\mathrm{t}}^ke^l\,,\label{solZe_bulk}
\end{align}
\end{subequations}
where $\omega=\frac{1}{3!}\varepsilon_{ijk}e^i\wedge e^j\wedge e^k$ is a volume form over $\Sigma$ because we work with non-degenerate frames. In the following we will use the notation
\[
f^{ij}:=\frac{1}{2}\left(\frac{e^j\wedge F^i}{\omega}\right)\,,\quad t^{ij}:=\frac{1}{2}\left(\frac{e^j\wedge De^i}{\omega}\right)\,.
\]
From the bulk equations \eqref{eq_bulk_1}, \eqref{eq_bulk_2} it follows that $f, t$ are symmetric matrices in the internal indices.
In terms of these objects we have
\[
F^i=f^{ij}\varepsilon_{jkl}e^k\wedge e^l\,,\quad De^i=t^{ij}\varepsilon_{jkl}e^k\wedge e^l\,,
\]
and we can write \eqref{solZ_bulk} in the form
\begin{subequations}\label{solZ_bulk_ft}
\begin{align}
Z_A^i&=DA_{\mathrm{t}}^i+2\varepsilon_{jkl}f^{ji}e_{\mathrm{t}}^ke^l\,,\label{solZA_bulk_ft}\\
Z_e^i&=De_{\mathrm{t}}^i-\varepsilon^i_{\phantom{i}lm}A_{\mathrm{t}}^le^m+2\varepsilon_{jkl}t^{ji}e_{\mathrm{t}}^ke^l\,,\label{solZe_bulk_ft}
\end{align}
\end{subequations}
whose interpretation is discussed in detail in \cite{HKP}. Here we just recall that introducing the vector field $\bm{\rho}\in\mathfrak{X}(\Sigma)$ defined in the whole of $\Sigma$ including its boundary by
\begin{align*}
    \imath_{\bm{\rho}}e^i=e_{\mathrm{t}}^i \ ,
\end{align*}
equations \eqref{solZA_bulk_ft} and \eqref{solZe_bulk_ft} can be rewritten in the form
\begin{subequations}\label{solZ_bulk_ft_Lie}
\begin{align}
Z_A^i&=\pounds_{\bm{\rho}}A^i+D(A_{\mathrm{t}}^i-\imath_{\bm{\rho}}A^i)\,,\label{solZA_bulk_Lie}\\
Z_e^i&=\pounds_{\bm{\rho}}e^i-\varepsilon^i_{\phantom{i}jk}(A_{\mathrm{t}}^j-\imath_{\bm{\rho}}A^j)e^k\,.\label{solZe_bulk_Lie}
\end{align}
\end{subequations}
The interpretation of \eqref{solZA_bulk_Lie} and \eqref{solZe_bulk_Lie} is well-known: the bulk dynamics corresponds to diffeomeorphisms and ``internal'' rotations defined by the arbitrary objects $\bm{\rho}$ and $A_{\mathrm{t}}^i-\imath_{\bm{\rho}}A^i$. In particular, the fact that some of these transformations are diffeomorphisms suggests that it will be convenient to restrict our model to a configuration space such that the vector field $\bm{\rho}$ is tangent to the boundary.
As we will see this can be achieved by imposing an extra boundary condition on the fields.

The boundary components of $\mathbb{Z}$ are even simpler to find as they are determined by the linear system of equations with constant coefficients \eqref{zbound}.
As we did above, it is convenient to rewrite these equations in the form
\begin{subequations}\label{Zbound_new}
\begin{align}
&2\alpha_3\jmath_\partial^*(Z_A^i-DA_{\mathrm{t}}^i)+\alpha_5\jmath_\partial^*(Z_e^i-De_{\mathrm{t}}^i+\varepsilon^i_{\phantom{i}jk}A_{\mathrm{t}}^je^j)
=-2\alpha_1\varepsilon^i_{\phantom{i}jk}\jmath_\partial^*(e_{\mathrm{t}}^je^k)\,,\label{Zbound1_new}\\
&\alpha_5\jmath_\partial^*(Z_A^i-DA_{\mathrm{t}}^i)+2\alpha_2\jmath_\partial^*(Z_e^i-De_{\mathrm{t}}^i+\varepsilon^i_{\phantom{i}jk}A_{\mathrm{t}}^je^j)
=-2\alpha_4\varepsilon^i_{\phantom{i}jk}\jmath_\partial^*(e_{\mathrm{t}}^je^k)\,,\label{Zbound2_new}
\end{align}
\end{subequations}
For a generic choice of $\alpha_1\,,\ldots\,,\alpha_5$ (i.e. when $\alpha_5^2-4\alpha_2\alpha_3\neq0$) the solutions to these equations are
\begin{subequations}\label{solZ_bound}
\begin{align}
&\jmath_\partial^*Z_A^i=\jmath_\partial^*(DA_{\mathrm{t}}^i)+2\alpha\jmath_\partial^*(\varepsilon^i_{\phantom{i}jk}e_{\mathrm{t}}^je^k)\,,\label{solZA_bound}\\
&\jmath_\partial^*Z_e^i=\jmath_\partial^*(De_{\mathrm{t}}^i-\varepsilon^i_{\phantom{i}jk}A_{\mathrm{t}}^je^k)+2\beta\jmath_\partial^*(\varepsilon^i_{\phantom{i}jk}e_{\mathrm{t}}^je^k)\,,\label{solZe_bound}
\end{align}
\end{subequations}
where
\[
\alpha:=\frac{2\alpha_1\alpha_2-\alpha_4\alpha_5}{\alpha_5^2-4\alpha_2\alpha_3}\,,\quad \beta:=\frac{2\alpha_3\alpha_4-\alpha_1\alpha_5}{\alpha_5^2-4\alpha_2\alpha_3}\,.
\]
By continuity the pullbacks of \eqref{solZ_bulk} (which are well defined because the components of the Hamiltonian vector field are differential forms) to the boundary must coincide with the values obtained in \eqref{solZ_bound}. This condition leads to the following additional set of boundary constraints:
\begin{subequations}\label{const_bound_new}
\begin{align}
&\jmath_\partial^*\Big(\big(f^{ji}-\alpha\delta^{ji}\big)\varepsilon_{jkl}e_{\mathrm{t}}^ke^l\Big)=0\,,\label{const1_bound_new}\\
&\jmath_\partial^*\Big(\big(t^{ji}-\beta\delta^{ji}\big)\varepsilon_{jkl}e_{\mathrm{t}}^ke^l\Big)=0\,.\label{const2_bound_new}
\end{align}
\end{subequations}
which can be generically written in the equivalent form
\begin{subequations}\label{const_bound_new2}
\begin{align}
&\jmath_\partial^*\Big(\big(\alpha_5 t^{ji}+2\alpha_3f^{ji}+\alpha_1\delta^{ji})\varepsilon_{jkl}e_{\mathrm{t}}^ke^l\Big)=0\,,\label{const1_bound_new2}\\
&\jmath_\partial^*\Big(\big(2\alpha_2 t^{ji}+\alpha_5f^{ji}+\alpha_4\delta^{ji})\varepsilon_{jkl}e_{\mathrm{t}}^ke^l\Big)=0\,.\label{const2_bound_new2}
\end{align}
\end{subequations}
At this point it is convenient to rewrite \eqref{secconstbdry} in terms of $f^{ij}$ and $t^{ij}$. These are
\begin{subequations}\label{const_bound_new3}
\begin{align}
&\jmath_\partial^*\left(\big(\alpha_5 t^{ij}+2\alpha_3f^{ij}+\alpha_1\delta^{ij}\big)\varepsilon_{jkl}e^k\wedge e^l\right)=0\,,\label{const1_bound_new3}\\
&\jmath_\partial^*\left(\big(2\alpha_2 t^{ij}+\alpha_5f^{ij}+\alpha_4\delta^{ij}\big)\varepsilon_{jkl}e^k\wedge e^l\right)=0\,.\label{const2_bound_new3}
\end{align}
\end{subequations}

Before we continue, there is a relatively fine point that must be considered with care. This concerns the possible extension of the bulk constraints to the boundary. The complete specification of the configuration space of the system requires a discussion of the smoothness conditions that the fields must satisfy. A simple way to proceed would be to demand as much regularity as needed to guarantee that all the expressions that appear in our analysis (for instance, the constraints) are well defined. This is in line with the traditional attitude in physics. However, in the presence of boundaries this has some consequences that have to be acknowledged and taken into account. Consider the bulk constraints. A relevant question regarding them is: Should they also hold at $\partial\Sigma$? In fact, intuitively one would expect the answer to be positive as a consequence of a simple and natural continuity requirement. The answer actually depends on the regularity conditions that we impose on the fields. For instance, let us take $\Sigma$ with a regular boundary such that it can be submersed in an open manifold $\widetilde{\Sigma}$. If we demand that all the basic fields---i.e. the variables defining our configuration space $(A_{\mathrm{t}}^i,A^i,e_{\mathrm{t}}^i,e^i)$---admit smooth extensions to $\widetilde{\Sigma}$ then, if they satisfy the constraints in the interior of $\Sigma$ they will also do so at $\partial\Sigma$ because $F_i$ and $De_i$ will be $C^\infty(\widetilde{\Sigma})$ and, hence, the constraints themselves when evaluated at these field configurations will also be smooth.
On the other hand it is conceivable that no inconsistencies arise if the bulk constraints are not required to hold at the boundary.
It may also happen that demanding consistency leads to conditions that are essentially equivalent to the extension property.
In the following we will work under the hypothesis that the bulk constraints hold at the boundary.

The tangency conditions for the bulk constraints \eqref{secconstbulk} are obtained by computing their directional derivatives along the field $\mathbb{Z}$. These are:
\begin{subequations}\label{tangency_bulk}
\begin{align}
&\varepsilon_{ijk}DZ_e^j\wedge e^k+\varepsilon_{ijk}\varepsilon^{jlm}Z_{Al}e^m\wedge e^k+\varepsilon_{ijk}De^j\wedge Z_e^k=0\label{tangency_bulk1}\ ,\\
&\varepsilon_{ijk}DZ_A^j\wedge e^k+\varepsilon_{ijk}F^j\wedge Z_e^k=0 \ .\label{tangency_bulk2}
\end{align}
\end{subequations}
It is possible to directly check that these equations hold for the values of $Z_A^i$ and $Z_e^i$ given in \eqref{solZ_bulk}. A better strategy is to consider them together with \eqref{Zbulk_new}. In fact, by computing the covariant differential $D$ of these two conditions we find
\begin{align}
&\varepsilon_{ijk}De^j\wedge Z_A^k+\varepsilon_{ijk}DZ_A^j\wedge e^k+\varepsilon_{ijk}F^j\wedge De_{\mathrm{t}}^k-A_{\mathrm{t}i}(F_j\wedge e^j)+F_i\wedge(A_{\mathrm{t}j}e^j)\label{alpha}\\
&\,\,+\varepsilon_{ijk}DA_{\mathrm{t}}^j\wedge De^k=0\,,\nonumber\\
&\varepsilon_{ijk}De^j\wedge Z_e^k-\varepsilon_{ijk}De^j\wedge De_{\mathrm{t}}^k+De^j\wedge(A_{\mathrm{t}i}e_j)-De^j\wedge(A_{\mathrm{t}j}e_i)+\varepsilon_{ijk}DZ_e^j\wedge e^k\label{beta}\\
&\,\,+(e^j\wedge F_i)e_{\mathrm{t}j}-(e^j\wedge F_j)e_{\mathrm{t}i}-e^j\wedge D(A_{\mathrm{t}i}e_j)+e^j\wedge D(A_{\mathrm{t}j}e_i)-\varepsilon_{ijk}De_{\mathrm{t}}^i\wedge De^k\nonumber\\
&\,\,-e_{\mathrm{t}}^j(F_i\wedge e_j)+e_{\mathrm{t}}^j(F_j\wedge e_i)=0 \ .\nonumber
\end{align}
By subtracting \eqref{tangency_bulk1} and \eqref{alpha} and using \eqref{solZA_bulk} and \eqref{Zbulk2_new} we get
\[
A_{\mathrm{t}j}(F_i\wedge e^j-F^j\wedge e_i)+\omega \varepsilon_{ijk}\varepsilon_{pqr}e_{\mathrm{t}}^q(f^{kp}t^{rj}-t^{kp}f^{rj}) \ ,
\]\
which can be seen to vanish by expanding $\varepsilon_{ijk}\varepsilon_{pqr}$ in terms of Kronecker deltas and using the secondary constraints in the form $f^{ij}=f^{ji}$, $t^{ij}=t^{ji}$. An analogous computation involving \eqref{tangency_bulk2} and \eqref{beta} shows that the second tangency condition in the bulk also holds.

The tangency conditions for the boundary constraints \eqref{secconstbdry} are
\begin{subequations}\label{tangency_bdry}
\begin{align}
& \jmath_\partial^*(\alpha_5 DZ_e^i+\alpha_5 \varepsilon^i_{\phantom{i}jk}Z_A^j\wedge e^k+2\alpha_3DZ_A^i+2\alpha_1\varepsilon_{ijk}Z_e^j\wedge e^k)=0\,,\label{tangency_bdry1}\\
& \jmath_\partial^*(2\alpha_2 DZ_e^i+2\alpha_2 \varepsilon^i_{\phantom{i}jk}Z_A^j\wedge e^k+\alpha_5DZ_A^i+2\alpha_4\varepsilon_{ijk}Z_e^j\wedge e^k)=0\,.\label{tangency_bdry2}
\end{align}
\end{subequations}
As we did before, instead of directly plugging the solutions for $\jmath_\partial^*Z_A^i$ and $\jmath_\partial^*Z_e^i$ into \eqref{solZ_bound}, it is better to compute the covariant differential $D$ of \eqref{zbound} (taking advantage of the fact that the pullback behaves well with respect to the exterior differential and the exterior product) to get
\begin{subequations}\label{Dpartials}
\begin{align}
&2\alpha_3\jmath_\partial^*(DZ_A^i-\varepsilon^i_{\phantom{i}jk}F^jA_{\mathrm{t}}^k)+\alpha_5\jmath_\partial^*\big(DZ_e^i-\varepsilon^i_{\phantom{i}jk}F^je_{\mathrm{t}}^k+\varepsilon^i_{\phantom{i}jk}D(A_{\mathrm{t}}^je^k)\big)\nonumber\\
&\,\,=-2\alpha_1\varepsilon^i_{\phantom{i}jk}\jmath_\partial^*\big(D(e_{\mathrm{t}}^je^k)\big)\,,\label{Dpartial3}\\
&\alpha_5\jmath_\partial^*(DZ_A^i-\varepsilon^i_{\phantom{i}jk}F^jA_{\mathrm{t}}^k)+2\alpha_2\jmath_\partial^*\big(DZ_e^i-\varepsilon^i_{\phantom{i}jk}F^je_{\mathrm{t}}^k+\varepsilon^i_{\phantom{i}jk}D(A_{\mathrm{t}}^je^k)\big)\nonumber\\
&\,\,=-2\alpha_4\varepsilon^i_{\phantom{i}jk}\jmath_\partial^*\big(D(e_{\mathrm{t}}^je^k)\big)\,.\label{Dpartial4}
\end{align}
\end{subequations}
Now, combining \eqref{Dpartial3} and \eqref{tangency_bdry1} together with the pullback to the boundary of \eqref{Zbulk} we immediately get
\[
\jmath_\partial^*\Big((\alpha_5\varepsilon^i_{\phantom{i}jk}De^j+2\alpha_3\varepsilon^i_{\phantom{i}jk}F^j-2\alpha_1e^i\wedge e_k)A_{\mathrm{t}}^k\Big)=0\,.
\]
Proceeding in an analogous way with \eqref{Dpartial4} and \eqref{tangency_bdry2} we find
\[
\jmath_\partial^*\Big((2\alpha_2\varepsilon^i_{\phantom{i}jk}De^j+\alpha_5\varepsilon^i_{\phantom{i}jk}F^j-2\alpha_4e^i\wedge e_k)A_{\mathrm{t}}^k\Big)=0\,.
\]
As we can see these two expressions vanish as a consequence of the boundary constraints \eqref{secconstbdry}.

We now discuss the constraints \eqref{const_bound_new2} and \eqref{const_bound_new3}.
Note that if $\bm{\rho}$ is tangent to the boundary, we can swap the interior product with the pullback. More precisely, denoting by $\overline{\bm{\rho}}$ the restriction of $\bm{\rho}$ to the boundary,
\begin{align*}
    \jmath_{\partial}^* \imath_{\bm{\rho}} \theta = \imath_{\overline{\bm{\rho}}} \jmath_{\partial}^*  \theta \ .
\end{align*}
If this happens, then
\begin{align*}
    \imath_{\overline{\bm{\rho}}} \jmath_{\partial}^* \left( \tensor{\varepsilon}{_i_j_k} e^j \wedge e^k \right) =  \jmath_{\partial}^* \imath_{\bm{\rho}} \left( \tensor{\varepsilon}{_i_j_k} e^j \wedge e^k \right) =  2 \jmath_{\partial}^* \left( \tensor{\varepsilon}{_i_j_k} e_t^j \wedge e^k \right) \ .
\end{align*}
This implies that the tangency of $\bm{\rho}$ to the boundary and \eqref{const_bound_new3} guarantee that the constraints \eqref{const_bound_new2} are satisfied.

We end this section by showing that the condition that $\bm{\rho}$ be tangent to $\partial\Sigma$ can be expressed in the form
\begin{equation}\label{extra_tangent}
\jmath_\partial^*(\varepsilon_{ijk}e_{\mathrm{t}}^ie^j\wedge e^k)=0\,,
\end{equation}
and checking the tangency of the vector field $\mathbb{Z}$ on the constraint submanifold of $\mathfrak{M}_0$ when this new condition is added.

On one hand we have (remember that $\partial\Sigma$ has dimension 2)
\[
0=\jmath_\partial^*(\varepsilon_{ijk}e^i\wedge e^j\wedge e^k)\Rightarrow\imath_{\overline{\bm{\rho}}}\jmath_\partial^*(\varepsilon_{ijk}e^i\wedge e^j\wedge e^k)=0\Rightarrow\jmath_\partial^*(\varepsilon_{ijk}e_{\mathrm{t}}^ie^j\wedge e^k)=0\,.
\]
Conversely, Let us suppose that $\jmath_\partial^*(\varepsilon_{ijk}e_{\mathrm{t}}^ie^j\wedge e^k)=0$, then we have that $\jmath_\partial^*\big(\imath_{\bm{\rho}}(\varepsilon_{ijk}e^i\wedge e^j\wedge e^k)\big)=0$. Let $X,Y\in\mathfrak{X}(\Sigma)$ be such that they are tangent to $\partial\Sigma$. Let us call $\overline{X}$ and $\overline{Y}$ their restrictions to $\partial\Sigma$. We have now $\imath_{\overline{X}}\imath_{\overline{Y}}\jmath_\partial^*(\imath_{\bm{\rho}}\omega)=0$, but this is the same as $0=\jmath_\partial^*(\imath_{X}\imath_{Y}\imath_{\bm{\rho}}\omega)=(\imath_{X}\imath_{Y}\imath_{\bm{\rho}}\omega)|\partial \Sigma$. As $\omega$ is a volume form this tells us that $\overline{X}$, $\overline{Y}$ and $\overline{\bm{\rho}}$ are linearly dependent on $\partial\Sigma$ and, hence, $\bm{\rho}$ is tangent to the boundary.

Finally let us look at the tangency condition for \eqref{extra_tangent}. Computing its Lie derivative along $\mathbb{Z}$ we find
\[
\jmath_\partial^*(\varepsilon_{ijk}Z_{e\mathrm{t}}^ie^j\wedge e^k+2\varepsilon_{ijk}e_{\mathrm{t}}^iZ_e^j\wedge e^k)=0\,.
\]
As the first term is proportional to $v_i Z_{e\mathrm{t}}^i$ ($v_i$ is constructed in the first lemma proved in appendix \ref{app_lemma}) and $v_i\neq0$ this equation can always be solved for $Z_{e\mathrm{t}}^i$, which guarantees the consistency of the tangency condition embodied by \eqref{extra_tangent}.

\medskip

\noindent \textbf{Remark 1.} As discussed, the consistency of the model depends on the vector field $\bm{\rho}$ being tangent to the boundary $\partial \Sigma$.
However, it is not necessary to impose this condition by hand.
Given the constraints \eqref{const_bound_new2}, \eqref{const_bound_new3}, the condition \eqref{extra_tangent} is automatically satisfied.
To see this, write
\begin{align*}
    C^i = \theta^{ij} \varepsilon_{jkl}e^k\wedge e^l \ ,
\end{align*}
where $\theta^{ij}$ is either $ \alpha_5 t^{ij}+2\alpha_3f^{ij}+\alpha_1\delta^{ij}$ or $2\alpha_2 t^{ij}+\alpha_5f^{ij}+\alpha_4\delta^{ij}$ so that
\begin{gather*}
\jmath_\partial^* C^i = 0 \ , \quad \jmath_\partial^* \imath_{\bm{\rho}} C^i = 0 \ ,
\end{gather*}
are equivalent to the constraints \eqref{const_bound_new3} and \eqref{const_bound_new2}, respectively.
Since $C^i$ is a 2-form whose pullback to the boundary vanishes, it can be written in a neighbourhood of $\partial \Sigma$ as $C^i = \tilde{C^i} \wedge n$, where $n$ is the normal to the boundary and $\tilde{C^i}$ is such that it depends on $e^i$, it is non-zero since $e^i$ is a non-degenerate triad, and its pullback does not vanish.
But then,
\begin{align*}
    \jmath_\partial^* \imath_{\bm{\rho}} C^i = - \jmath_\partial^*\imath_{\bm{\rho}}n \  \jmath_\partial^*\tilde{C^i} \ ,
\end{align*}
so that if condition \eqref{const_bound_new2} is satisfied, necessarily $\jmath_\partial^*i_{\bm{\rho}}n = 0$, hence $\bm{\rho}$ is tangent to $\partial \Sigma$.
This means that it is not necessary to impose this as an independent condition as it is built into the model to begin with.

\medskip

\noindent \textbf{Remark 2.} We assumed all the fields to smoothly extend into the boundary. However, one might wonder in which way does relaxing this assumption affect the results.
By using a similar argument to the one in the previous remark, it can be seen how even in this case, the model still enforces $\bm{\rho}$ to be tangent to the boundary.

\medskip

\noindent \textbf{Remark 3.}
Consider $\barbelow{e^i} \in \Omega^1(M)$ the part of $e^i$ tangent to each leaf $\Sigma$ and choose the volume form $\omega = \tensor{\varepsilon}{_i_j_k} dt \wedge \barbelow{e^i} \wedge \barbelow{e^j} \wedge \barbelow{e^k}$.
Since
\begin{align*}
    \tensor{\varepsilon}{_i_j_k} e^i \wedge e^j \wedge e^k = \tensor{\varepsilon}{_i_j_k} \barbelow{e^i} \wedge \barbelow{e^j} \wedge \barbelow{e^k} + 3 \tensor{\varepsilon}{_i_j_k} e_{\mathrm{t}}^i dt \wedge \barbelow{e^j} \wedge \barbelow{e^k} \ ,
\end{align*}
one easily computes
\begin{align*}
    \imath_{\mathbf{u}} \barbelow{e^i} =& \left( \frac{\barbelow{e^i}  \wedge \tensor{\varepsilon}{_j_k_l} e^j \wedge e^k \wedge e^l  }{\omega} \right) = \left( \frac{\barbelow{e^i}  \wedge \tensor{\varepsilon}{_j_k_l} \barbelow{e^j} \wedge \barbelow{e^k} \wedge \barbelow{e^l} }{\omega} \right) + 3\left( \frac{\barbelow{e^i}  \wedge \tensor{\varepsilon}{_j_k_l} e_{\mathrm{t}}^j dt \wedge \barbelow{e^k} \wedge \barbelow{e^l} }{\omega} \right) = \\
    &= -\frac{1}{2} \tensor{\varepsilon}{_j_k_l} \tensor{\varepsilon}{^i^k^l} e_{\mathrm{t}}^j  \omega =  -e_{\mathrm{t}}^i = -\rho^i \ ,\\
    \imath_{\mathbf{u}} \barbelow{dt} =& \left( \frac{dt  \wedge \tensor{\varepsilon}{_j_k_l} e^j \wedge e^k \wedge e^l  }{\omega} \right) = \left( \frac{dt \wedge \tensor{\varepsilon}{_j_k_l} \barbelow{e^j} \wedge \barbelow{e^k} \wedge \barbelow{e^l} }{\omega} \right) + 3\left( \frac{dt  \wedge \tensor{\varepsilon}{_j_k_l} e_{\mathrm{t}}^j dt \wedge \barbelow{e^k} \wedge \barbelow{e^l} }{\omega} \right) = \\
    &=1 \ .
\end{align*}
Hence we can decompose $\bm{u} = \partial_t - \bm{\rho}$.
Of course, $\bm{u}$ is a vector density and had we chosen a different volume form to define it, we would have obtained the same result up to a product with a non-vanishing function.
Since we are only interested in its direction, this suffices.
We know that $\bm{\rho}$ is tangent to the boundary $\partial \Sigma$, in particular it is also tangent to the boundary $\partial M$. On the other hand, $\partial_t$ is tangent to $\partial M$ by construction. Hence, $\bm{u}$ is also tangent to $\partial M$.

\begin{center}

\begin{tikzpicture}[scale=1]

\draw[rounded corners=35pt, dashed](0,0)--(0,4);
\draw[rounded corners=35pt, dashed](4.2,0)--(4.2,4);

\draw[-stealth](2,1.82)--(2,2.8);
\draw[-stealth](2,1.82)--(2.5,3);
\draw[-stealth](2,1.82)--(3.2,1.78);

\draw [rounded corners=35pt,dashed] (0,4) arc (260:280:12.12cm);
\draw [rounded corners=35pt] (0,2) arc (260:280:12.12cm);
\draw [rounded corners=35pt,dashed] (0,0) arc (260:280:12.12cm);

\node (a) at (0.5,3.6) {$\partial_{L}M$};
\node (b) at (1.8,2.4) {$\partial_{t}$};
\node (c) at (2.5,2.5) {$\bm{u}$};
\node (d) at (2.7,1.6) {$\bm{\rho}$};
\end{tikzpicture}

\medskip

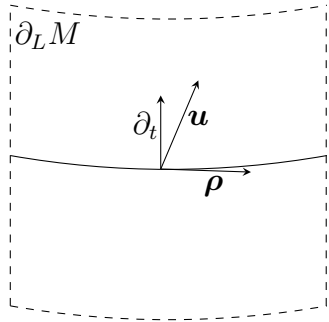
\textbf{Figure 2.} Tangency of the vector fields $\partial_{t}$, \textbf{u} and \textbf{$\rho$} to the boundary.
\end{center}

%
%
\section{Conclusions and comments}{\label{sec_conclusions}}


In this paper we have discussed three different methods to obtain the Hamiltonian formulation for the generalised Husain-Kucha\v{r}-Pontryagin action in a 4-dimensional manifold with boundary. A relevant feature of our analysis is that we have been able to deal with the boundary rigorously and in doing so, we have arrived at some key insights into the model. By appropriately choosing the coupling constants $\alpha_i$ in the bulk Lagrangian,
it is possible to get, as boundary contributions, the 3-dimensional Euclidean Einstein equations with an
arbitrary cosmological constant.

While boundaries may introduce new terms in the Lagrangian or Hamiltonian and require new constraints at the boundary, their effect is more profound than that. In particular, boundaries shape the configuration space and greatly affect the integrability conditions of the theory as discussed at the beginning of section \ref{sec_HKP} and throughout section \ref{subsec_tangency}. Note that integrability can be studied in the methods we have used, while in the covariant phase space treatments such as  \cite{Freidel:2020xyx,Freidel:2020svx,Margalef-Bentabol:2020teu,G:2021xvv,G:2021eey} (see also [38, 39] for 3-dimensional general relativity), which are focused on conserved charges and symmetries, integrability issues are not regarded.

Although all of the approaches used in section \ref{sec_HKP} to find the dynamics have produced equivalent results, it is interesting to compare them.

First, notice that all the methods give the same results because the Lagrangian is \emph{linear} in the velocities. An interesting consequence of this is the fact that the dynamics is fully contained in the configuration space $Q$ in the sense that, for instance, canonical momenta play no role. As a result, the formulations in $TQ$ and $T^*Q$, which share their base space, are equivalent.

Perhaps the most prominent difference between the approaches discussed here is the fact that the analysis presented in section \ref{subsec_GNH} is made in the cotangent space $T^*Q$ while the treatment in sections \ref{subsec_Pepin}, \ref{subsec_field_eqs} uses the tangent space $TQ$. Although they are equivalent, one might prefer one formulation over the others depending on the desired application, for instance, working in $T^*Q$ could be useful for quantization.

It is important to point out that if one chooses to work in $TQ$, one must additionally impose the second order condition \eqref{eq_second_order_condition} for the solutions to be equivalent to those coming from the Euler-Lagrange equations.
In $T^*Q$ such condition is not needed.

The geometric constraint algorithm (on which sections \ref{subsec_GNH}, \ref{subsec_Pepin} are both based) arose as a geometrized version of Dirac's method and is an improvement of it in the sense that it replaces Poisson bracket computations (that can be tricky, for instance if the fields are defined in a manifold with boundary) with geometric considerations. The field equations approach used in section \ref{subsec_field_eqs} manages to bypass many of the computations of the geometric constraint algorithm making it the fastest and least computationally involved of the three. In some known examples (for instance for general relativity written in terms of the Hilbert-Palatini or Holst actions) this is, by far, the best way to arrive at the Hamiltonian formulation \cite{Holstnos,Concise}. This is so because, in this case, it is crucially possible to simplify some of the field equations before applying the procedure that we have used here.

\appendix

\section{Some useful mathematical results}\label{app_lemma}

\noindent\textbf{Lemma 1}

\medskip

\noindent\emph{Let $\Sigma$ be a 3-dimensional manifold with parallelizable boundary $\partial\Sigma$. If $\varepsilon_{ijk}e^i\wedge e^j\wedge e^k$ is a volume form in $\Sigma$ (including at $\partial\Sigma$) then there exist $v^i\in C^\infty(\partial\Sigma)$ with $\delta^{ij}v_iv_j=1$ such that $\jmath_\partial^*(\varepsilon_{ijk} e^j\wedge e^k)=v_i\cdot\mathsf{area}$ where $\mathsf{area}$ is a volume form in $\partial\Sigma$.}

\bigskip

\begin{proof}
Let us take three everywhere linearly independent vector fields $X,Y,Z\in\mathfrak{X}(\Sigma)$ such that $X$ and $Y$ are tangent to the boundary and always different from zero there (remember that $\partial\Sigma$ is parallelizable) and $Z$ everywhere transverse to the boundary. At every point of $\partial\Sigma$ we have then
\[
\jmath_\partial^*\big(\imath_Z\imath_Y\imath_X (\varepsilon_{ijk}e^i\wedge e^j\wedge e^k)\big)\neq0\,.
\]
Now this implies
\[
\varepsilon_{ijk}\jmath_\partial^*(\imath_Ze^i)\jmath_\partial^*(\imath_Xe^j)\jmath_\partial^*(\imath_Ye^k)\neq0\,,
\]
but
\[
\varepsilon_{ijk}\jmath_\partial^*(\imath_Xe^j)\jmath_\partial^*(\imath_Ye^k)=\varepsilon_{ijk}\imath_{\tilde{X}}\imath_{\tilde{Y}}(\jmath_\partial^*e^j\wedge\jmath_\partial^*e^k) \ ,
\]
where $\tilde{X}$ and $\tilde{Y}$ are vector the fields at the boundary obtained by restricting $X$ and $Y$. Hence we conclude that $\varepsilon_{ijk}\jmath_\partial^*(e^j\wedge e^k)$ is different from zero everywhere on $\partial\Sigma$. This means that, for a given volume form $\mathsf{area}$ we have
\[
\varepsilon_{ijk}\jmath_\partial^*(e^j\wedge e^k)=\left(\frac{\varepsilon_{ijk}\jmath_\partial^*(e^j\wedge e^k)}{\mathsf{area}}\right)\mathsf{area} \ ,
\]
with $\displaystyle \left(\frac{\varepsilon_{ijk}\jmath_\partial^*(e^j\wedge e^k)}{\mathsf{area}}\right)$ everywhere different from zero. By normalizing it we get the desired $v_i$, which is unique modulo a sign.
\end{proof}

\bigskip

\noindent\textbf{Lemma 2}

\medskip

\noindent\emph{Let $M$ be a four dimensional parallelizable manifold, $e^i\in\Omega^1(M)$, with $i=1,2,3$, be linearly independent 1-forms and $S^i\in\Omega^1(M)$ be another three 1-forms.  Then $\varepsilon^i_{\phantom{i}jk}e^j\wedge S^k=0$ implies $S^i=0$.}

\medskip

\begin{proof}
Let us complete $e^i$ with a 1-form $e^0$ linearly independent with the $e^i$.
As $M$ is orientable $\varepsilon_{ijk}e^i\wedge e^j\wedge e^k\wedge e^0$ will be a volume form.
Let us expand now $S^i=S^i_{\phantom{i}j}e^j+S^i_{\phantom{i}0} e^0$. The condition  $\varepsilon^i_{\phantom{i}jk}e^j\wedge S^k=0$ then becomes
\begin{equation}\label{eclemma2}
\varepsilon^i_{\phantom{i}jk}e^j\wedge(S^k_{\phantom{k}l}e^l+S^k_{\phantom{i}0} e^0)=0\,.
\end{equation}
Let $\bm{u} \in \mathfrak{X}(M)$ be a vector field such that $e^0(\bm{u}) \neq 0$ and $e^i(\bm{u})=0$, then, by taking the interior product with $\bm{u}$ we find
\[
S^k_{\phantom{i}0}\varepsilon^i_{\phantom{i}jk}e^j=0\Rightarrow S^k_{\phantom{i}0}\varepsilon^i_{\phantom{i}jk}=0\Rightarrow S^k_{\phantom{i}0}=0\,.
\]
Also, by taking the exterior product with $e^m\wedge e^0$ and using the fact that $\varepsilon_{ijk}e^i\wedge e^j\wedge e^k\wedge e^0$ is a volume form we immediately get
\[
S^k_{\phantom{k}l}\varepsilon^{i}_{\phantom{i}jk}\varepsilon^{jlm}=0\Rightarrow\delta^{im}S-S^{mi}=0\Rightarrow S^{mi}=0\,.
\]
Hence we conclude that $S^i = 0$.
\end{proof}

%
%
\section*{Acknowledgments}
This work has been supported by the Spanish Ministerio de Ciencia Innovaci\'on y Uni\-ver\-si\-da\-des-Agencia Estatal de Investigaci\'on FIS2017-84440-C2-2-P and PID2020-116567GB-C22 grants. E.J.S. Villase\~nor is supported by the Madrid Government (Comunidad de Madrid-Spain) under the Multiannual Agreement with UC3M in the line of Excellence of University Professors (EPUC3M23), and in the context of the V PRICIT (Regional Programme of Research and Technological Innovation).

\end{document}